%% file: secure-shuffler-survey.tex
\documentclass[acmsmall,authorversion,nonacm]{acmart} 

\acmJournal{CSUR}

\usepackage{subcaption}
\usepackage{multirow}
\usepackage{pifont}
\newcommand{\cmark}{\hfil\ding{51}}%
\newcommand{\xmark}{\hfil\ding{55}}%

\newcommand\pt[1]{{\left(#1\right)}}
\DeclareMathOperator{\enc}{Enc}
\DeclareMathOperator{\dec}{Dec}
\DeclareMathOperator{\keygen}{KeyGen}

\usepackage{xcolor}
\usepackage{soul}

\begin{document}

\title[How to Securely Shuffle?]{How to Securely Shuffle? A survey about Secure Shufflers for privacy-preserving computations}


\author{Marc Damie}
\email{m.f.d.damie@utwente.nl}
\orcid{0000-0002-9484-4460}
\affiliation{%
    \institution{University of Twente}
    \city{Enschede}
    \country{The Netherlands}
    \and
    \institution{Inria}
    \city{Lille}
    \country{France}
    \thanks{This work was supported by the Netherlands Organization for Scientific Research (NWO) in the context of the SHARE project [CS.011],  the ANR project ANR-20-CE23-0013 PMR, and by the Horizon Europe project HE TRUMPET}
}

\author{Florian Hahn}
\orcid{0000-0003-4049-5354}
\affiliation{%
    \institution{University of Twente}
    \city{Enschede}
    \country{The Netherlands}}

\author{Andreas Peter}
\orcid{0000-0003-2929-5001}
\affiliation{%
    \institution{Carl von Ossietzky Universität Oldenburg}
    \city{Oldenburg}
    \country{Germany}
}

\author{Jan Ramon}
\orcid{0000-0002-0558-7176} 
\affiliation{%
    \institution{Inria}
    \city{Lille}
    \country{France}}

\renewcommand{\shortauthors}{Damie et al.}

\begin{abstract}
    Ishai et al. (FOCS'06) introduced secure shuffling as an efficient building block for private data aggregation.
    Recently, the field of differential privacy has revived interest in secure shufflers by highlighting the privacy amplification they can provide in various computations.
    Although several works argue for the utility of secure shufflers, they often treat them as black boxes; overlooking the practical vulnerabilities and performance trade-offs of existing implementations.
    This leaves a central question open: what makes a good secure shuffler?

    This survey addresses that question by identifying, categorizing, and comparing 26 secure protocols that realize the necessary shuffling functionality.
    To enable a meaningful comparison, we adapt and unify existing security definitions into a consistent set of properties.
    We also present an overview of privacy-preserving technologies that rely on secure shufflers, offer practical guidelines for selecting appropriate protocols, and outline promising directions for future work.
\end{abstract}



\maketitle

\section{Introduction}

Since the emergence of privacy regulations such as GDPR or CCPA, there has been an increased interest in privacy-preserving technologies, in particular in applications involving massive consumer data.
The literature has described many approaches to ``enhance'' privacy in privacy-preserving computations.
Among all the approaches, secure shufflers play an important role as they offer at the same time: accurate results and high privacy guarantees.
A secure shuffler is a cryptographic protocol (involving one or multiple servers) that takes in messages from a group of senders, shuffles them, and forwards them to a receiver without revealing the origin of each message.

This line of work originates from \citet{ishai_cryptography_2006} that proposed, in 2006, secure shufflers to perform privacy-preserving data aggregation.
Recent works in Differential Privacy (DP) \cite{bittau_prochlo_2017,balle_improved_2019,balle_privacy_2019,cheu_distributed_2019,erlingsson_amplification_2019,cheu_differential_2021} revived this interest in secure shufflers by formalizing the ``privacy amplification'' they provide.
Based on DP analysis, shuffling-based private computations offer privacy-utility trade-off close to private computations performed by a trusted third party (i.e., a trusted party offering the optimal trade-off in DP).
Thus, secure shufflers have been presented as an efficient solution to build privacy-preserving computations without requiring a trusted party.

In addition to the privacy amplification in DP, secure shufflers have found applications in numerous fields from privacy-preserving software monitoring \cite{bittau_prochlo_2017} to Federated Learning (FL) \cite{konecny_federated_2016,konecny_federated_2017}; a decentralized ML paradigm in which data owners collaboratively without sharing their personal data.
In these applications, the shufflers enable to compute efficiently and privately a form of data aggregation without revealing the individual values.
Recently, Wang et al. \cite{wang_beyond_2025} broadened the scope of applications as they highlighted the potential of secure shufflers in ``private individual computations'' (e.g., taxi hailing).

Most papers introducing such shuffling-based computations (e.g., \cite{ghazi_scalable_2019,cheu_distributed_2019}) use an abstract shuffler rather than a concrete implementation.
An advantage of considering an abstract shuffler is to focus on the statistical properties, which are relevant for the applications, and plug in later a concrete shuffler implementation.
However, not all secure shufflers are equally appropriate for all applications, as shuffler and application papers make their own explicit or implicit security assumptions, which must be compatible.

Understanding properties of secure shuffler implementations has several benefits.
First, it allows for a fairer comparison between private computation algorithms.
For example, existing papers do not tell whether shuffling-based data aggregation is more scalable than masking-based data aggregation \cite{mansouri_sok_2023}, since most papers on shuffling-based computations do not detail their shuffler implementation.
Second, it contributes to a better formalization of the concept of ``secure shuffler''.
The literature relies on a vague definition leaving some blind spots; e.g., several papers \cite{ghazi_scalable_2019,cheu_distributed_2019,kairouz_advances_2021} present mix networks as an example of secure shufflers, without mentioning the possible de-anonymization and censorship attacks \cite{shmatikov_timing_2006,sun_raptor_2015}.
Third, it could lead to more consensus on what technique is appropriate in which use cases.
For example, Ghazi et al. \cite{ghazi_scalable_2019} present mix nets as ``highly scalable'' while Gordon et al. \cite{gordon_spreading_2022} point at the same shufflers as too expensive, which may seem contradictory without a deeper understanding of the context.

Implementing an ideal shuffler (as defined in the theory) is not straightforward.
\citet{bagdasaryan_towards_2022} pointed out the \textbf{need for a thorough understanding of secure shuffler implementations} as they highlighted that only one DP paper \cite{bittau_prochlo_2017} detailed their implementation.
Our survey comes to fill this gap.

\paragraph{Our contributions}
\begin{enumerate} 
    \item Identify and \textbf{compare 26 existing secure protocols} fulfilling the minimum requirements of a secure shuffler.
    \item Adapt their original security definitions to a \textbf{uniform set of security properties} that supports our comparison.
    \item Survey the applications of secure shufflers in privacy-preserving computations.
    \item Identify several \textbf{open problems} (e.g., robustness to data poisoning).
    \item Introduce \textbf{guidelines} to choose a shuffler based on the properties of a privacy-preserving application.
\end{enumerate}

Our paper has the following structure:
\begin{itemize}
    \item Section \ref{sec:background} provides background knowledge on cryptography and differential privacy.
    \item Section \ref{sec:def} introduces four security properties to \emph{define secure shufflers}.
    \item Section \ref{sec:comparison} \emph{compares the existing methods} based on their scalability and security (using the properties defined in Section \ref{sec:def}). Tables \ref{tab:summary_efficiency}, \ref{tab:summary_security}, and \ref{tab:summary_msg_size} summarize the main comparison points.
    \item Section \ref{sec:applications} presents \emph{the applications} of secure shufflers.
    \item Section \ref{sec:how_to_choose} presents \emph{guidelines for choosing a secure shuffler} and demonstrate them.
    \item Section \ref{sec:takeaways} presents several general \emph{perspectives} on the literature.
    \item Section \ref{sec:conclusion} concludes the paper and lists \emph{promising future works}.
\end{itemize}

\input{sec/background.tex}

\input{sec/definitions.tex}

\input{sec/comparison.tex}

\input{sec/applications.tex}

\input{sec/how_to_choose.tex}

\input{sec/takeaways.tex}

\section{Conclusion and future directions}
\label{sec:conclusion}
This survey highlighted the interest and properties of secure shufflers.
These secure protocols can support privacy-preserving computations involving thousands or even millions of data owners.
However, guaranteeing the security properties (conditioning the privacy guarantees) requires some attention.
We identified and compared six categories of secure shufflers.
We provided guidelines and examples to help an ML researcher choose the adequate secure shuffler depending on the use case.
Hence, secure shufflers are a promising line of work for privacy-preserving computations, but some gaps remain in the literature:

\textbf{Optimization of anonymous broadcasters for the ``few-receivers/many-senders'' setup.}
Many anonymization systems, motivated by applications such as anonymous whistleblowing, are optimized for setups with few senders and many receiver.
However, secure shuffling usually involved an opposite setup many senders and very few receivers (i.e., the analyzer).
Thus, leveraging the properties of this setup may improve multi-server secure shuffling.

\textbf{Linear complexity for 1-server secure shuffling.}
Recent works \cite{bittau_prochlo_2017,shi_non-interactive_2021,bunz_non-interactive_2021} approached linear complexity (i.e., $O(n)$ communications and computations) for 1-server secure shuffling, the optimal complexity for 1-server shuffling.

Prochlo has linear complexity but is vulnerable to side-channel attacks.
Recent \emph{secure} TEE-based solutions \cite{ngai_distributed_2024,gu_efficient_2023} avoided these attacks, but have a complexity of $O(n\log n)$.
Finally, NIDAR verifies fewer security properties than the rest of the solutions and only has subquadratic costs.

Hence, research is still necessary to achieve a linear complexity with satisfying security properties (or to provide an impossibility result for this complexity).

\textbf{Robustness to data poisoning} Shuffling-based privacy-preserving computations are particularly sensitive to data poisoning attacks (see Subsection \ref{subsec:malicious-sender}).
The anonymization provided makes it hard (or even impossible) to detect ``poisonous'' inputs from the shuffle output.
Attackers can easily provide maliciously crafted input biasing the result (e.g., adding a backdoor to a model in FL).

The literature describes no mitigation compatible with shuffling-based protocols.
Thus, research is needed to build new mitigations specifically for shuffling-based privacy-preserving computations.

\textbf{Formalization of the ``imperfect'' shuffle DP model.}
Several works \cite{kwon_atom_2017,bunz_non-interactive_2021,langowski_trellis_2023,zhou_power_2022,balle_amplification_2023} proposed relaxed anonymity definitions to conceive more scalable secure shufflers.
While some works \cite{zhou_power_2022,balle_amplification_2023} adapted the shuffle DP model to their ``imperfect'' shuffling, no DP results exist for the relaxation of Langowski et al. \cite{langowski_trellis_2023}.
Thus, further theoretical work is still needed to formalize these relaxations, and present DP results for shufflers with a relaxed anonymity definition.

\textbf{Computational shuffle DP.}
Subsection \ref{subsec:comp-shuffle-dp} explained why only computational Shuffle DP is practical.
While Mironov et al. \cite{mironov_computational_2009} showed that computational DP can reach better accuracy guarantees than information-theoretic DP, no work has yet studied this relaxation in Shuffle DP.
Hence, analogous investigations for Shuffle DP would be necessary to take advantage of our impossibility result and build more accurate computations.

\textbf{New types of shuffling-based private computation.}
While most existing applications are using secure shufflers to perform a form of data aggregation, recent works \cite{gascon_computationally_2024,wang_beyond_2025} recently opened the way to other privacy-sensitive applications such as information retrieval or taxi hailing could be enabled by secure shufflers.
Thus, further research into new shuffling-based private computations is a promising direction.
The concept of ``private individual computation'' introduced by \citet{wang_beyond_2025} represents a valuable framework to structure such research effort.


\bibliographystyle{ACM-Reference-Format}
\bibliography{ref}


\end{document}

%% file: sec/background.tex
\section{Background}
\label{sec:background}

\subsection{Terminology}
In privacy-preserving computations, a group of data owners $\{P_1, \ldots P_N\}$ (each owning a private value $x_i$) wants to compute a function $f(x_1, \dots f_N)$.
However, they want to reveal as little information as possible.
Our survey focuses on privacy-preserving computations during which the data owners are supported by a ``secure shuffler'' to preserve privacy.

A secure shuffler $\mathcal{S}$ is a set $\{S_1 \ldots S_M\}$ of $M$ shuffling servers.
Each data owner can send none, one, or more messages to $\mathcal{S}$.
Let $n$ be the total number of messages sent to $\mathcal{S}$.
The shuffler $\mathcal{S}$ outputs the shuffled message set $\{m_1 \ldots m_n\}$; all outputs usually transmitted to a party (e.g., a bulletin board or a data analyser) who will further process it.
In particular, the set $\{m_1 \ldots m_n\}$ contains the messages in random order; i.e., does not reveal the sender of any of the messages.
Section \ref{sec:def} further formalizes this concept.

\subsection{Differential Privacy}
Dwork \cite{dwork_differential_2006} introduced differential privacy (DP) to characterize the privacy leakage of any computation with public output:

\begin{definition}[Differential Privacy]
  A randomized algorithm $\mathcal{M}: \mathbb{X}^N \rightarrow \mathbb{Z}$ is $\pt{\epsilon, \delta}$-differentially private if for any datasets $D$ and $D'$ that differ on a single element and for all subsets $S$ of $\mathbb{Z}$:
  $$
    \Pr[\mathcal{M}(D) \in S] \le e^\epsilon \cdot \Pr[\mathcal{M}(D') \in S] + \delta
  $$
  \label{def:central-dp}
\end{definition}

The space $\mathbb{X}$ corresponds to the secret value space (i.e., each data owner $P_i$ owns a private value $x_i \in \mathbb{X}$).
The parameter $\epsilon \in [0, \infty)$ controls the privacy loss: smaller $\epsilon$ implies more privacy.
To ensure strong privacy, $\epsilon$ is usually chosen to be below 1 \cite{abadi_deep_2016}.
The parameter $\delta \in [0,1)$ relaxes the definition and allows for rare cases where the privacy loss would be higher than $\epsilon$.
To ensure strong privacy, the probability $\delta$ should be below $1/|D|$ (with $D\in \mathbb{X}^N$, a dataset) \cite{abadi_deep_2016}.

To reduce the $\epsilon$, DP algorithms add random noise to the computation.
This added noise creates a privacy-accuracy trade-off (sometimes called privacy-utility in the literature).
The goal of DP works is to optimize this trade-off.

Definition \ref{def:central-dp} is called ``central'' DP because it was originally intended for setups with a trusted curator collecting all data centrally and publishing a privacy-preserving output.

Central DP provides the optimal trade-off because the trusted curator can simply add a small noise at the end of the computation.
However, it requires a trusted curator to process the data, which is not necessarily possible.

\textbf{Local DP.}
If there is no central trusted curator, local differential privacy (LDP) is an option that requires adding more noise but privatizes the data before it leaves the data owner.
In particular, LDP introduces a local randomizer $\mathcal{R}: \mathbb{X}\rightarrow \mathbb{Y}$.

\begin{definition}
  The local randomizer $\mathcal{R}$ is $\pt{\epsilon, \delta}$-local differentially private if for any two data samples $x,x'\in \mathbb{X}$ and all subset $S$ of $\mathbb{Y}$:
  $$
    \Pr[\mathcal{R}(x) \in S] \le e^\epsilon \cdot \Pr[\mathcal{R}(x') \in S] + \delta
  $$
  \label{def:local-dp}
\end{definition}

If $\mathcal{R}$ is $\pt{\epsilon, \delta}$-local differentially private, then $\mathcal{M}(x_1\dots x_n) = (\mathcal{R}(x_1)\dots\mathcal{R}(x_n))$ is $\pt{\epsilon, \delta}$-DP.

The local randomizer used in LDP induces noisier results than central DP, but LDP makes no security or trust assumptions; less trust then requires higher noises.

Several papers have proposed strategies to ``amplify the privacy''; i.e., reducing the amount of noise needed to reach a certain privacy level.
These solutions include iterative protocols \cite{feldman_privacy_2018}, decentralization \cite{cyffers_privacy_2021}, random check-ins\cite{balle_privacy_2020} and \emph{shuffling} \cite{erlingsson_amplification_2019} (detailed in Subsection \ref{subsec:shuffle_dp}).

\subsection{Cryptographic building blocks}
\subsubsection{Encryption schemes}
Encryption schemes are defined by three functions $(\keygen, \enc, \dec)$ respectively responsible for encryption and decryption.
The key generation algorithm $\keygen(1^\lambda)$ is a randomized algorithm taking as input a security parameter $\lambda$ and outputting one key (for symmetric encryption) and two keys (for public-key encryption).

The security parameter $\lambda$ is a fundamental concept in cryptography that quantifies the security level in a cryptographic scheme; determining notably the size of the keys.
A PPT (Probabilistic Polynomial-Time) adversary is a computationally bounded attacker whose runtime is polynomial in $\lambda$.
Secure cryptographic schemes must ensure that the probability of a PPT adversary to break it is negligible in $\lambda$; i.e., it decreases faster than any inverse polynomial function of $\lambda$.

The encryption function $\enc$ takes as input a key $k_{\enc}$ and a plaintext $m$: $\enc(k_{\enc}, m) = c$.
The decryption function $\dec$ takes as input a key $k_{\dec}$ and a ciphertext $c$: $\dec(k_{\dec}, c) = m$.
When an agent possesses the correct keys, we have $\dec(k_{\dec}, \enc(k_{\enc}, m)) = m$.

In symmetric encryption, we use the same key for encryption and decryption: $k_{\enc} = k_{\dec}$.
In public-key encryption, the keys are different: $pk$ (called public key) for encryption and $sk$ (called secret key) for decryption.



\subsubsection{Secret Sharing}
Shares are a set of values that, once gathered, reveal a secret value.
Individually, the shares leak no information about the secret value.
A secret value $x$ is considered shared when several parties own a share; $P_j$ owning the share $x_j$.
Several secret sharing schemes exist, e.g., additive shares are numbers drawn randomly subject to the constraint that their sum (in a finite field) reveals the shared value: $\sum_j x_j = x$.
Recent works also enable more complex forms of secret sharing, such protocols to share a function \cite{boyle_function_2015}.

\subsubsection{Multi-party computation (MPC)}
MPC \cite{evans_pragmatic_2018} is a set of techniques enabling several parties to compute a function over secret inputs.
Many MPC protocols rely notably on secret sharing schemes.
The main bottleneck of MPC is the number of parties: while they provide high security, state-of-the-art protocols scale quadratically with the number of shareholders (i.e., they would not be applied for more than a handful of parties).

\subsubsection{Zero-knowledge proofs (ZKP)}
ZKPs \cite{goldreich_definitions_1994} are techniques to prove properties about ciphertexts (e.g., that two encrypted values are equal or that an encrypted value is in a given range) without disclosing anything else.
A ZKP involves two agents: the prover (who provides the proof) and the verifier (who verifies its correctness).

\subsubsection{Trusted Execution Environment (TEE)}
\label{subsec:tee-definition}

A trusted execution environment (TEE) \cite{sabt_trusted_2015} (sometimes called ``secure enclave'') is a secure environment in which confidentiality and integrity for the processed data and code are ensured.
Even the hardware owner cannot read or modify the data processed in the TEE.
Anyone can audit the code executed in the environment, and the hardware design ensures that the data processed inside remains confidential.

They require no trust other than in the manufacturer.
Many major chip manufacturers have already designed such environments; e.g., Intel SGX \cite{intel_software_2015}.

While they are attractive due to their minimal requirement for significant infrastructure changes, they are not without their limitations.
Indeed, TEEs are theoretically secure, but their practical security can be compromised through side-channel attacks.
Subsection \ref{subsec:tee-strong-assumptions} details these security shortcomings.

\subsection{Basic shuffling-based data aggregation}
\label{subsec:ikos}
Ishai et al. \cite{ishai_cryptography_2006} introduced the first shuffling-based private computation.
In particular, they presented a shuffling-based protocol to aggregate private values.
Their work is the seminal work that inspired most recent works on shuffling-based private computations \cite{balle_improved_2019,wang_beyond_2025}.

The protocol begins with each data owner splitting their private value into several additive shares of the private value.
The shares are individually sent (i.e., one message per share) to a secure shuffler that publishes its shuffling output.
One can sum all the shuffled shares to obtain the \emph{exact} sum of the initial private values.

\citet{ishai_cryptography_2006} showed that each data owner only needs to send a logarithmic number of shares in order to have shares indistinguishable from perfect randomness.
To reconstruct the input secret values based on the shuffled random shares, an adversary would need to compute an exponential number of share combinations.
The shuffling combined with the share randomization then guarantees that an attacker cannot realistically harm the input privacy, but allows to efficiently perform a data aggregation while preserving input privacy.

%% file: sec/definitions.tex
\section{Secure shuffler definition}
\label{sec:def}

This section formalizes the security properties desirable for a secure shuffler.

\subsection{Threat models}
Threat models characterize agents or behaviors that threaten a desirable security property of a protocol (e.g., correctness or privacy).
When defining a secure protocol, it is necessary to specify a threat model against which the security properties are guaranteed.
In other words, the desirable property of a protocol holds as long as no adversary is more powerful (i.e., has more knowledge or computational power) than what is defined in the threat model.
This subsection reminds the terminology commonly used in security literature \cite{evans_pragmatic_2018}. 

Threat modelling distinguishes three types of agents: \emph{honest agents} (following the protocol), \emph{semi-honest agents} (who follow the protocol but try to infer some private information passively), and \emph{malicious agents} (who do not follow the protocol and can perform actions to disrupt the protocol and gain additional information).
A malicious agent can notably be a corrupted device or server.
While semi-honest agents may compromise only the privacy of a computation, malicious agents also threaten the correctness of the computation.

Secure protocols typically guarantee data privacy against a proportion of malicious/semi-honest agents.
For instance, Clarion \cite{eskandarian_clarion_2022} is secure as long as one out of $M$ servers (involved in the protocol) is honest.
On the other hand, RPM \cite{lu_rpm_2023} is secure as long as there is a majority of honest servers involved in the protocol.

Two agents sharing their information (outside the protocol) are referred to as  ``colluding'' agents.
Threat models usually consider a unique adversary composed of all the corrupted parties (to be as conservative as possible in the security analysis).

While most secure protocols restrict their threat model to the agents actively involved in the protocol, anonymous communication protocols can also be the target of traffic analysis attacks \cite{murdoch_low-cost_2005}.
Some works then include in their threat model a ``global adversary'' able to capture the network traffic of the agents.
A ``global adversary'' is typically a state or an internet service provider.

\subsection{Shuffler definition}
Our starting point is the definition of a shuffler $\mathcal{S}$.
Based on papers studying shuffling-based DP \cite{cheu_distributed_2019,cheu_differential_2021}, we propose the following naive definition:
\begin{definition}
  A \textbf{Shuffler} $\mathcal{S}: \{0,1\}^{n\times L} \rightarrow \{0,1\}^{n\times L} $ takes as input a collection of messages $(m_1 \dots m_n)$ (for any $m_i \in \{0,1\}^{L}$) and outputs $(\widetilde{m}_1 \dots \widetilde{m}_n)$, with $(\widetilde{m}_1 \dots \widetilde{m}_n) = \pi(m_1 \dots m_n)$ and $\pi$ a uniformly randomly drawn permutation on the set $\{1 \ldots n\}$.
  \label{def:shuffler}
\end{definition}

This naive definition offers an intuitive understanding of the problem at hand, but it is too informal.
On the one hand, it considers the secure shuffler as a unique and well-identified entity, while secure shuffling protocols commonly require the participation of multiple parties, notably the data owners.
On the other hand, it does not reuse classic security frameworks necessary to provide formal security proofs.

Indeed, several secure shuffling techniques involve local preprocessing at the sending parties, e.g., mix networks (presented in Subsection \ref{subsec:comp_mix_net}) require performing multilayer encryption.
While Definition \ref{def:shuffler} is an appropriate definition for a shuffling function executed locally, it does not capture the reality of a secure shuffling protocol, which typically involve multiple parties and algorithms.
Thus, we propose to refine this definition, and characterize a shuffler using a pair of randomized algorithms $\pt{\mathcal{P}, \mathcal{M}}$:

\begin{itemize}
  \item The local preprocessor $\mathcal{P}: \{0,1\}^L \rightarrow \mathcal{Y}$ is executed by the sender on their message $m_i$.
  \item The mixer $\mathcal{M}: \mathcal{Y}^{n} \rightarrow \{0,1\}^{L\times n}$ takes as input the pre-processed messages and outputs messages (i.e., a permutation of the input messages). The mixer is composed of $M$ servers (where $M\ge 1$).
\end{itemize}

\begin{figure}
  \centering
  \includegraphics[width=0.3\linewidth]{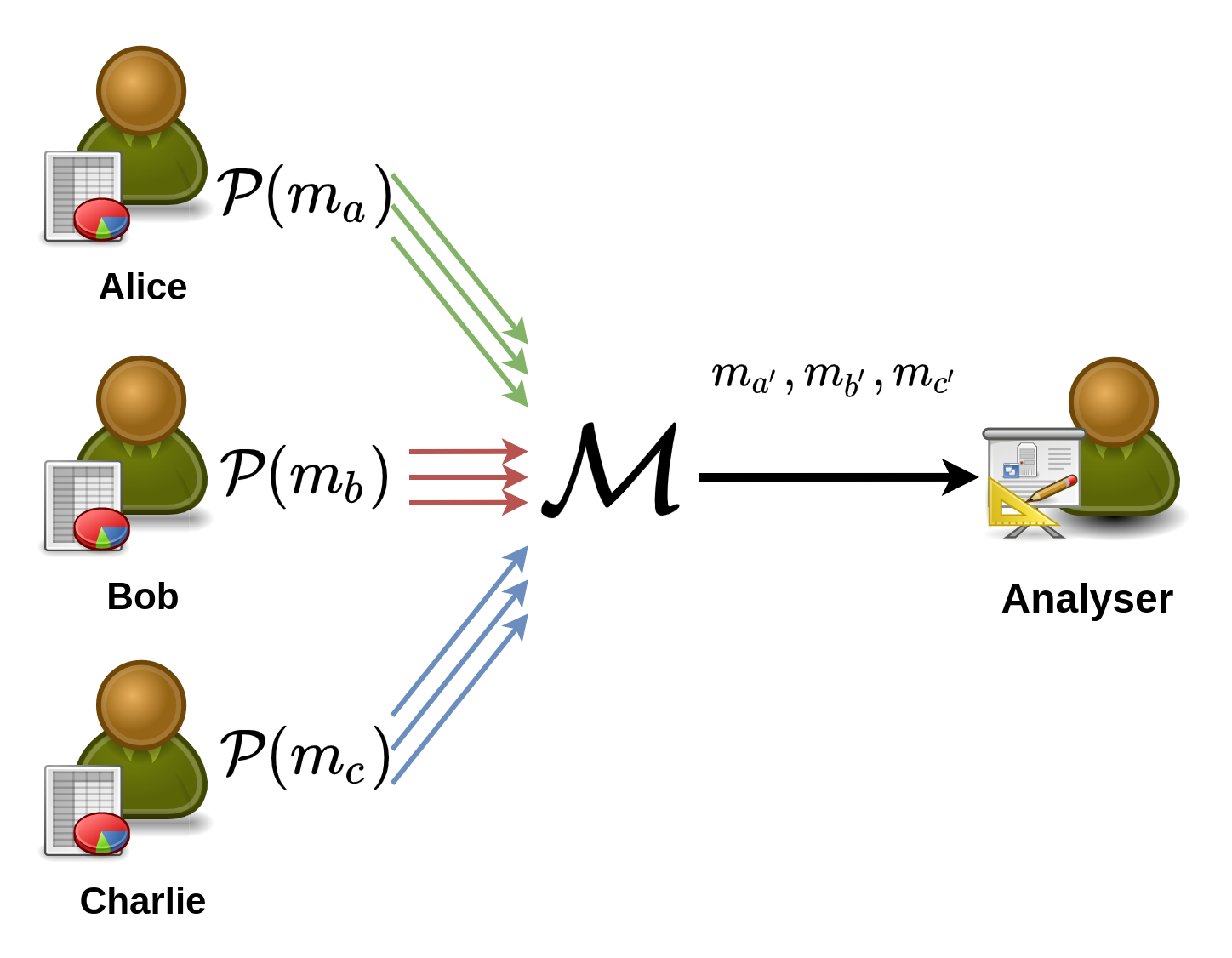}
  \caption{Abstract shuffler design: (1) The senders pre-process their messages using $\mathcal{P}$, (2) They send their preprocessed messages to the mixer $\mathcal{M}$, (3) The mixer outputs a permutation of the messages, and (4) an analyser can use the shuffled messages in a privacy-preserving computation.}
  \label{fig:abstract-shuffler}
  \Description[Schema of an abstract shuffler]{Schema of an abstract shuffler: three data owners preprocess their private data and send them to a mixer M; transmitting the shuffled output to an analyzer.}
\end{figure}

The local preprocessor $\mathcal{P}$ implicitly includes the communication between the senders and the mixer.
The shuffle can then be expressed as $\mathcal{S}(m_1, \dots, m_n) = \mathcal{M}(\mathcal{P}(m_1),\dots, \mathcal{P}(m_n))$ (as represented in Fig. \ref{fig:abstract-shuffler}).
The mixer $\mathcal{M}$ is the component that is usually considered when referring to a secure shuffler.
For simplicity, we call ``shuffling servers'' the servers executing the mixer $\mathcal{M}$.
For example, a shuffler based on a single trusted server is defined by $(\mathcal{P}_T,\mathcal{M}_T)$ with $\mathcal{P}_T(m) = m$ and $\mathcal{M}_T(m_1, \dots, m_n) = \pi(m_1,\dots m_n)$; $\mathcal{M}_T$ is executed by the trusted server.

We have now a definition of ``shuffler'' but not yet of ``\emph{secure} shuffler''.
Let us start from the following \textbf{informal} definition:
\begin{definition}
  A shuffler $\mathcal{S}$ is \textbf{secure} against a threat model $\mathcal{A}$ if, for all messages $m_1 \dots m_n \in \{0,1\}^{L}$, there exists a permutation $\pi$ such that $\mathcal{S}(m_1, \dots, m_n) = \pi(m_1, \dots, m_n)$ where no adversary fitting the threat model $\mathcal{A}$ can infer anything about $\pi$.
  \label{def:intuitive-secure-shuffler}
\end{definition}

This informal definition fits what the literature \cite{kairouz_advances_2021} \emph{implicitly} considers as a ``secure shuffler'' (also referred to as ``oblivious shuffler'' in \cite{bittau_prochlo_2017,gordon_spreading_2022}).
However, this definition is imprecise.
Thus, we propose to formalize this notion using four new security properties to characterize a shuffler $\mathcal{S}$: \textbf{Anonymity, Correctness, Sender disruption resistance, Server disruption resistance}.

\subsection{Security properties}
\label{subsec:sec_prop}
\subsubsection{Anonymity}
\label{sec:anonymity}
This is the minimum requirement for shuffling-based private computations.
More precisely, one is usually interested in sender anonymity because only the origin of the messages must remain anonymous.
It is even the core concern of shuffler application paper that give more importance to the concept of ``anonymous messaging'' than to shuffling itself \cite{ishai_cryptography_2006,cheu_distributed_2019,ghazi_pure_2020}.


\paragraph{Absence of standard formalization}
While the first anonymization techniques \cite{chaum1981untraceable,chaum_dining_1988} appeared in the 1980s, the first anonymity formalization paper \cite{feigenbaum_model_2007} was only published in 2007.
Thus, most anonymization systems (up until now) have been built upon informal anonymity definitions.
For example, even recent constructions such as Clarion \cite{eskandarian_clarion_2022} prove the security of their MPC protocol, but do not prove the anonymity itself; \cite{eskandarian_clarion_2022} does not even cite any formalization framework.

\citet{pfitzmann_terminology_2010} proposed a recognized terminology for all the words related to anonymity \cite{lu_survey_2019}.
These definitions are informal but are a commonly accepted starting point.
They give the following definition for sender anonymity:

\begin{definition}
  A sender $P$ is anonymous with regard to sending if and only if $P$ is not distinguishable within the set of potential senders.
  \label{def:anonymity-pfitzmann}
\end{definition}

Finding a formal definition for anonymity is a complex problem because anonymity is context-dependent.
Several papers \cite{backes_anoa_2013,gelernter_limits_2013,kuhn_privacy_2019} proposed formal anonymity definitions, but they remain imperfect.
For example, some of them exclude some de-anonymization attacks (e.g., traffic analysis attacks in \cite{backes_anoa_2013}), which creates a blind spot in the theory that struggles to fit the reality of anonymity.
Furthermore, \citet{lu_survey_2019} surveyed these works on anonymity formalizations and concluded that the formalization proposed ``is still a theoretic approach and lacks a practical verification.''
Hence, anonymity formalization is still an open problem for the security community that is still looking for a standard and general solution.

\paragraph{Theoretic definition}
Our survey does not intend to solve this long-running formalization problem.
Instead, we propose a theoretic definition of the sender anonymity and then highlight five factors that can degrade this anonymity in practice.
Our theoretic definition formalizes Definition \ref{def:anonymity-pfitzmann} using standard security notions, especially indistinguishability.

Let $\mathcal{V}(\cdot)$ be the view/knowledge of an agent (e.g., a sender or server involved in the protocol), and let $\mathcal{H}$ be the set of honest senders. We propose the following theoretic anonymity definition;

\begin{definition}{(Anonymity)}
  A shuffler $\mathcal{S}$ provides \textbf{anonymity against a threat model} $\mathcal{A}$ if, for any message $m \in \mathcal{S}(m_1,\dots, m_n)$ (with $m_i\sim \mathcal{U}(\{0,1\}^L), \forall i\in\{1\dots n\}$), any PPT adversary $A$ fitting the threat model $\mathcal{A}$, and any honest senders $P,P'\in\mathcal{H}$, we have:
  \begin{align*}
    \lvert\Pr[m \text{ sent by } P \vert \mathcal{V}(A)] - & \Pr[m \text{ sent by } P' \vert \mathcal{V}(A)]\rvert<f(\lambda),       \\
                                                                  & \text{ where $f$ is a negligible function and $\lambda$ a security parameter.}
  \end{align*}
  \label{def:anonymity_vs}
\end{definition}

This definition means that an adversary can distinguish honest senders only with negligible probability.
Suppose all the messages are unique, the probability of $m$ being sent by the honest party $P$ (i.e., $\Pr[m \text{ sent by } P \vert \mathcal{V}(A)]$) has two possible values: $0$ if the adversary controls the sender of $m$, and $\frac{1}{\lvert\mathcal{H}\rvert}$, otherwise.
If the messages are not all unique, we can have $q$ honest senders who sent the same message $m$; the probability of a (honest) $P$ being the sender of $m$ is $\frac{q}{\lvert\mathcal{H}\rvert}$.

\paragraph{Degrading factors} Five factors can degrade this theoretic definition in practice:
\begin{itemize}
  \item \emph{Message content}: A message is anonymous if it is impossible to infer its sender from its content (e.g., the message ``I am Alice'' is straightforward to de-anonymize). This requirement is present in Definition \ref{def:anonymity_vs} via the uniform distribution of the messages. This de-anonymization factor is not related to the shuffler implementation. In other words, the protocol using the shuffler must ensure that the distribution of the messages sent does not reveal their origin.
  \item \emph{Metadata leakage}: In all surveyed techniques but one, a passive traffic observer can infer the size and number of messages sent by each sender. If all messages have a different size, one can link the shuffled messages to their sender by comparing the size of the sent and shuffled messages. Subsection \ref{subsec:ultimate_anon} explains why avoiding this leakage would induce an impractical cost and discusses why it is tolerable in privacy-preserving computations.
  \item \emph{Censorship}: in some systems, the adversary can censor some senders. In these cases, this decreases the quality of the anonymization because it reduces the number of potential senders. The second security property of this section (i.e., correctness) guarantees that censorship is not possible at the shuffler level. However, an adversary may still attempt a network-level attack (e.g., blocking the sender's network traffic).
  \item \emph{Uniformity of the anonymization}: for optimization purposes, some shufflers may result in non-uniform distributions over the permutation $\pi$; i.e., the $f(\lambda)$ of Definition \ref{def:anonymity_vs} is not negligible but still small enough to be useful in a number of practical applications. Subsection \ref{subsec:imperfect_shuffle} discusses this relaxation.
  \item \emph{De-anonymization attacks}: Subsection \ref{subsec:attacks} presents a taxonomy of attacks against secure shufflers. Some attacks are agnostic to the shuffler implementation, while others are implementation-specific.
\end{itemize}

\paragraph{Comparison with surveys on anonymous communications}
Our survey focuses on sender anonymity as it is the form of anonymity required by secure shufflers.
However, some anonymous communication systems provide other forms of anonymity such as one-to-one anonymous communications \cite{van_den_hooff_vuvuzela_2015,eskandarian_express_2021} (i.e., the two communicating agents know the identity of their correspondent, but there is a guarantee that no third party knows these agents are communicating).
This is useful in a number of applications but incompatible with sender anonymity since the receiver knows the sender's identity.
Hence, such systems cannot be used as secure shufflers.
Among all anonymous communication systems, only ``anonymous broadcasters'' satisfy the secure shuffling needs.

Over the years, several works including \cite{edman2009anonymity,sasy_sok_2023,shirali_survey_2022,shirazi2018survey} have surveyed anonymous communication systems.
As \citet{sasy_sok_2023} is the most recent and complete survey, we will focus our discussion on it, but our remarks hold for all surveys on anonymous communications.
While both \cite{sasy_sok_2023} and our work gravitates around the notion of anonymity, they provide different perspectives because they do not focus on the same functionality.
On the one hand, they focus on anonymous communications.
On the other hand, we focus on secure shuffling.

Since we focus on different functionalities, our work relies on a different set of definitions.
For example, the concept of correctness (defined in the next subsubsection) is not studied in \cite{sasy_sok_2023}.
Indeed, correctness is an important notion in privacy-preserving computations as we want to avoid a bias in the output results.
The anonymous communications literature is much more interested in the notion of censorship resistance (which is implied by correctness but is not equivalent).

This scope difference has a visible consequence: while some solutions are covered in both works, some solutions are only present in \cite{sasy_sok_2023} (e.g., one-to-one anonymous communications), and others are only present in our work (e.g., TEE-based shufflers).

A core contribution of our work is to identify secure protocols able to implement a secure shuffler, even though these protocols were not initially designed for it.
While some anonymous communications systems can implement this functionality, many other secure protocols provide the sender anonymity necessary to secure shuffling.
For example, Section \ref{sec:comparison} highlights ZKP protocols initially used for electronic voting \cite{adida_helios_2008} and  TEE solutions imagined for secure contact discovery \cite{ngai_distributed_2024}.

\subsubsection{Correctness}
Next to anonymity, a good shuffler should also satisfy correctness.
This property covers various security definitions present in existing security papers.
On the one hand, secure shufflers based on MPC (e.g., RPM \cite{lu_rpm_2023}) prove a standard security property called ``correctness'' (i.e., the MPC protocol outputs the expected output).
On the other hand, recent anonymous communication systems (e.g., Blinder \cite{abraham_blinder_2020}) often guarantee censorship resistance.
This second security guarantee is essential, but it does not prevent all the threats covered by the ``correctness.''
Taking inspiration from MPC, we propose the following explicit definition for shuffling correctness:

\begin{definition}{(Correctness)}
  A shuffler $\mathcal{S}$ is \textbf{correct} against threat model $\mathcal{A}$, if, in the presence of any PPT adversary $A\in\mathcal{A}$, for all uninterrupted protocol runs:
  \begin{align*}
    \Pr[\exists \pi,\mathcal{S}(m_1,\ldots,m_n) & = \pi(m_1,\ldots,m_n)]  \ge 1 - f(\lambda),                                    \\
                                                       & \text{ where $f$ is a negligible function and $\lambda$ a security parameter.}
  \end{align*}
  \label{def:correctness}
\end{definition}

This property implies that the shuffle output contains all the input messages (and only those) because an adversary has only a negligible chance to alter the result.

\paragraph{Importance}
Shuffler correctness prevents three types of attacks (executed by malicious shuffling servers): censorship, message tampering, and message injection.
In privacy-preserving computations, these attacks would impact all the users because it biases the final result.
Indeed, a malicious server may try to censor a specific type of sender to bias an ML model or a statistical estimation.
Hence, shuffle correctness is necessary for the overall computation correctness.

Moreover, some theoretical models (such as the Shuffle DP) rely on an abstract shuffler model fitting Definition \ref{def:shuffler}, which implicitly requires correctness.

\paragraph{Formal secure shuffler definition}
Using the previous two security properties, we can formalize the secure shuffler definition as follows:
\begin{definition}{(Secure Shuffler)}
  A shuffler $\mathcal{S}$ is \textbf{secure} against a threat model $\mathcal{A}$ if it \textbf{guarantees anonymity and correctness} against $\mathcal{A}$.
  \label{def:secure-shuffler}
\end{definition}

\subsubsection{Sender disruption resistance}
Even if we have a formal secure shuffler definition, additional security definitions can address other practical issues not covered by the intuitive definition.

Several ML works studied the resilience of distributed training protocols to malicious behaviors \cite{chen_distributed_2017,alistarh_byzantine_2018}.
These works conceive protocols that succeed even in the presence of malicious or faulty agents.
Protocol disruption is also possible in secure shuffling and is critical when thousands or millions of senders are involved.
Controlling the honesty of millions of senders is too complex, so a form of disruption resistance is necessary.
We propose the following definition regarding the resistance to malicious senders:

\begin{definition}{(Sender disruption resistance)}
  A shuffler $\mathcal{S}$ is \textbf{resistant to sender disruption} if malicious senders cannot force protocol abortion by sending ill-formed data or not sending data.
  \label{def:sender_disrupt}
\end{definition}

This property provides tolerance to drop-outs (i.e., the disconnections of senders before the protocol completion), which drew much attention in the secure aggregation literature \cite{bell_secure_2020,sabater_accurate_2021,bell_acorn_2022}.
Secure aggregation protocols such as \cite{bell_secure_2020,bell_acorn_2022} only tolerate a fixed number of drop-outs, while sender disruption-resistant shufflers tolerate an unbounded number of drop-outs.
Sender disruption resistance is stronger than drop-out tolerance since it also covers the reception of ill-formed data.

Definition \ref{def:sender_disrupt} does not limit the number of malicious senders: a disruption-resistant shuffler can always produce an output.
However, publishing the output is not desirable if the output size is too small to guarantee satisfying anonymity.
Subsection \ref{subsec:malicious-sender} further discusses this issue.

Finally, Definition \ref{def:sender_disrupt} specifies that the disruption can only be caused by missing or ill-formed messages to exclude Distributed Denial-of-Service (DDoS) attacks (i.e., flooding a server with a massive amount of requests to make it unavailable).
These attacks (or any other network-level attack) are not specific to shufflers and can be mitigated using network security techniques \cite{mirkovic_taxonomy_2004}.

\subsubsection{Server disruption resistance}
Finally, we can consider server disruption resistance:

\begin{definition}{(Server disruption resistance)}
  A shuffler $\mathcal{S}=(\mathcal{P}, \mathcal{M})$ is \textbf{resistant to server disruption} if a malicious shuffling server (i.e., involved in $\mathcal{M}$) cannot force protocol abortion by sending ill-formed data or not sending data.
  \label{def:server_disrupt}
\end{definition}

This property is called ``robustness'' in MPC protocols \cite{abraham_blinder_2020}.
One may see it as an extra property because the number of servers is small enough to maintain a blocklist of untrustworthy servers.
As for sender disruption resistance, server disruption resistance brings resilience to drop-outs.

In summary, (sender or server) disruption resistance is not necessary to define a secure shuffler but preferable for real-world applications.

\subsection{Taxonomy of de-anonymization attacks against secure shufflers}
\label{subsec:attacks}
This subsection provides a taxonomy of de-anonymization attacks and discusses mitigation.

\subsubsection{Intersection attacks}
These attacks \cite{wolinsky_hang_2013,danezis_statistical_2005} rely on the statistical analysis that can be made from several shuffle outputs to de-anonymize messages.
Indeed, if a data owner participates in $k$ shuffles, computing the intersection of the round results might leak some information about this data owner.
These attacks against shufflers are specific to the use case and the data being shuffled.
For example, the intersection attack is useless if the shuffled messages are secret shares, as in Ishai et al. \cite{ishai_cryptography_2006} protocol, but the intersection may de-anonymize the messages if we shuffle cleartext data.

Wolinsky et al. \cite{wolinsky_hang_2013} propose a generic \textit{Buddy} system compliant with any anonymization system to mitigate these attacks.
The general idea is to ensure a large set of possible senders for a given message stream, so the intersection of multiple shuffles cannot lead to a de-anonymization.
No result is available regarding the scalability of this generic mitigation technique for applications to privacy-preserving computations.

\subsubsection{Trickle and flooding attacks}
Serjantov et al. \cite{goos_trickle_2003} describe two general attacks (grouped under the term ``blending attacks''): trickle attacks and flooding attacks.

Their idea is to force the shuffle output only to contain one honest message.
The attacker can then easily infer the source of this communication.
In trickle attacks, the adversary censors messages from everyone but one user.
In flooding attacks, the adversary uses colluding malicious data owner to ``flood'' the batch with malicious messages and one honest message.

The trickle attack is only possible on a system insecure against censorship (i.e., with no correctness guarantee as defined in Section \ref{sec:def}).
The flooding attack is only possible if a malicious server colludes with data owners and can be mitigated using an authentication scheme, so a single malicious user cannot flood a batch alone.
Some secure shufflers (e.g., mix nets) must use anonymous authentication systems \cite{chase_algebraic_2014} to mitigate these attacks.

\subsubsection{Traffic analysis attacks}
These attacks \cite{murdoch_low-cost_2005,sun_raptor_2015,arp_torben_2015} have been extensively documented for mix networks: a global adversary (e.g., a state agency) observes the packet transmission between the agents to trace the path of the messages.
From the beginning, anonymization networks such as Tor were presented as possibly insecure against global attackers.

However, we can describe two simple countermeasures to mitigate these attacks.
First, we can distribute the agents in several countries.
This practice decreases the chance of having a global adversary able to observe all agents.
Second, increasing the traffic makes message tracing harder.

Only mix networks is sensitive to this attack, while other secure shufflers are secure by design.

\subsubsection{Shuffler-specific attacks}
The previous paragraphs cover the main attacks affecting all secure shufflers.
However, new secure solutions often bring new attack vectors, specific to the proposed technique.

For example, mix nets are sensitive to replay attacks (i.e., retransmitting a message received during a previous shuffle to infer its sender using the intersection of the shuffle outputs).

On the other hand, TEE-based solutions can be sensitive to side-channel attacks (i.e., attacks exploiting leakage from the protocol implementation, but not the protocol itself) in which the memory access patterns help de-anonymize a shuffle of some constructions \cite{sasy_fast_2022}.

%% file: sec/comparison.tex
\section{Comparison of the existing secure shufflers}
\label{sec:comparison}

This section identifies and compares 26 existing protocols implementing a secure shuffler.
We distinguish six categories of secure shufflers (based on design similarities): mix networks, verifiable shufflers, DC nets, MPC-based shufflers, TEE-based shufflers, and other shufflers.
We summarize the comparison in Tables \ref{tab:summary_efficiency} and \ref{tab:summary_security}, which respectively analyze scalability and security.

\subsection{Methodology}
We adopted two approaches to identify relevant works able to implement a secure shuffler.
First, we performed a survey on Google Scholar using different combinations of keyword including ``secure shuffle,'' ``anonymous broadcaster,'' ``verifiable shuffler,'' ``mix network,'' ``anonymous communications,'' ``shuffle differential privacy,'' and ``secret shuffle.''
This first step provided a preliminary list of secure shufflers.
Second, we used Google Scholar to find additional secure shufflers in papers that cited these constructions.
This second step enables to identify follow-up works that may propose more unconventional solutions to securely shuffle.
For example, Bell et al. \cite{bell_secure_2020} introduced a secure-aggregation-based shuffler, while the main focus of their paper is secure aggregation.

To classify a work as relevant, we need the proposed protocol to \textit{at least} provide a form a sender anonymity (as defined in Section \ref{sec:anonymity}).
This requirement is satisfied by many secure protocols from anonymous communications to MPC-based shuffling.
Thus, our surveying process required examining works from various security topics including anonymous communications, secure enclaves, and electronic voting.
Through this process, we identified 26 secure shufflers that we have grouped into 6 categories based on their design similarities.

\begin{table}
    \footnotesize
    \begin{center}
        \renewcommand{\arraystretch}{1.2}
        \caption{Scalability comparison of the secure shuffling solutions}
        \label{tab:summary_efficiency}
        \begin{tabular}{|p{0.18\linewidth}|p{0.10\linewidth}|p{0.10\linewidth}|p{0.14\linewidth}|p{0.14\linewidth}|c|p{0.06\linewidth}|}
            \hline
            \textit{Technique}                                                                  & \textit{Per-Sender comm.} & \textit{Per-Sender comp}. & \textit{Per-Server comm.}                  & \textit{Per-Server comp.}      & $M$      & \textit{Open Source} \\\hline
            Baseline: Trusted shuffler                                                          & $O(1)$                    & $O(1)$                    & $O(n)$                                     & $O(n)$                         & 1        & \xmark               \\\hline
            \multicolumn{7}{|c|}{\textbf{Mix networks} ($k$ = path length)}                                                                                                                                                                                             \\\hline
            Fixed-route \cite{chaum1981untraceable,danezis_mixminion_2003,freedman_tarzan_2002} & $O(1)$                    & $O(k)$                    & $O(n)$                                     & $O(n)$                         & $ \ge 1$ & \xmark               \\\hline
            Free-route \cite{chaum1981untraceable,dingledine2004tor,chen_hornet_2015}           & $O(1)$                    & $O(k)$                    & $O(nk/M)$                                  & $O(nk/M)$                      & $\gg 1$  & \xmark               \\\hline
            Advanced \cite{chaum_cmix_2017,piotrowska_loopix_2017}                              & $O(1)$                    & $O(k)$                    & $O(nk/M)$                                  & $O(nk/M)$                      & $\gg 1$  & \cmark               \\\hline
            Trellis \cite{langowski_trellis_2023}                                               & $O(1)$                    & $O(k)$                    & $O(nk/M)$                                  & $O(nk/M)$                      & $\gg 1$  & \cmark               \\\hline

            TORaaS   \cite{hohenberger_anonize_2014}                                            & $O(1)$                    & $O(k)$                    & $O(n)$                                     & $O(n)$                         & 1        & \cmark               \\\hline

            \hline
            \multicolumn{7}{|c|}{\textbf{Verifiable shufflers}}                                                                                                                                                                                                         \\\hline
            Classic \cite{furukawa_efficient_2001,neff_verifiable_2001,bayer_efficient_2012}    & $O(M)$                    & $O(1)$                    & $O(M\sqrt{n} + Mn)$                        & $O(n \log n + Mn + M\sqrt{n})$ & $>1$     & \cmark               \\\hline
            Riffle \cite{kwon_riffle_2016}                                                      & $O(1)$                    & $O(M)$                    & $O(n)$                                     & $O(n)$                         & $>2$     & \cmark               \\\hline
            Atom \cite{kwon_atom_2017} ($k$: group size, $\alpha=\frac{k}{M}$)                  & $O(k)$                    & $O(1)$                    & $O(n\cdot \alpha + k\sqrt{n\cdot \alpha})$ & $O(\alpha \log \alpha)$        & $\gg 1$  & \cmark               \\\hline
            Lattice-based \cite{aranha_verifiable_2023}                                         & $O(M)$                    & $O(1)$                    & $O(Mn)$                                    & $O(Mn)$                        & $>1$     & \xmark               \\\hline

            \hline
            \multicolumn{7}{|c|}{\textbf{DC networks}}                                                                                                                                                                                                                  \\\hline
            Dissent \cite{corrigan-gibbs_dissent_2010}                                          & $O(Mn)$                   & $O(Mn)$                   & $O(n^2 + nM)$                              & $O(Mn)$                        & $>1$     & \cmark               \\\hline
            Riposte \cite{corrigan-gibbs_riposte_2015}                                          & $O(\sqrt{n})$             & $O(\sqrt{n})$             & $O(n\sqrt{n})$                             & $O(n^2)$                       & $3$      & \cmark               \\\hline
            Spectrum \cite{newman_spectrum_2022}                                                & $O(\log n )$              & $O(\log n)$               & $O(n\log n)$                               & $O(n^2)$                       & 2        & \cmark               \\\hline
            Sabre \cite{vadapalli_sabre_2022}                                                   & $O(\log n )$              & $O(\log n)$               & $O(n\log n)$                               & $O(n^2)$                       & 3        & \xmark               \\\hline

            \hline
            \multicolumn{7}{|c|}{\textbf{MPC-based shufflers}}                                                                                                                                                                                                          \\\hline
            AsynchroMix \cite{lu_honeybadgermpc_2019}                                           & $O(M)$                    & $O(M)$                    & $O(M\cdot n \cdot log^2 n)$                & $O(M\cdot n \cdot log^2 n)$    & $\ge 4$  & \cmark               \\\hline
            PowerMix \cite{lu_honeybadgermpc_2019}                                              & $O(M)$                    & $O(M)$                    & $O(n^2)$                                   & $O(n^3)$                       & $\ge 4$  & \cmark               \\\hline
            Blinder  \cite{abraham_blinder_2020}                                                & $O(M\sqrt{n})$            & $O(M\sqrt{n})$            & $O(n\sqrt{n})$                             & $O(n^2)$                       & $\ge 5$  & \cmark               \\\hline
            Clarion  \cite{eskandarian_clarion_2022}                                            & $O(M)$                    & $O(M)$                    & $O(Mn)$                                    & $O(Mn)$                        & $\ge 3$  & \cmark               \\\hline
            RPM  \cite{lu_rpm_2023}                                                             & $O(M)$                    & $O(M)$                    & $O(Mn\sqrt{n})$                            & $O(Mn^2)$                      & $\ge 3$  & \cmark               \\\hline
            Ruffle  \cite{a_ruffle_2023}                                                        & $O(M)$                    & $O(M)$                    & $O(Mn)$                                    & $O(Mn)$                        & $= 3$    & \cmark               \\\hline

            \hline
            \multicolumn{7}{|c|}{\textbf{TEE-based shufflers}}
            \\\hline
            Prochlo \cite{bittau_prochlo_2017}                                                  & $O(1)$                    & $O(1)$                    & $O(n)$                                     & $O(n)$                         & 1        & \cmark               \\\hline
            ORShuffle \cite{sasy_fast_2022}                                                     & $O(1)$                    & $O(1)$                    & $O(n)$                                     & $O(n\log^2 n)$                 & 1        & \cmark               \\\hline
            Butterfly Shuffle \cite{gu_efficient_2023}                                          & $O(1)$                    & $O(1)$                    & $O(n)$                                     & $O(n\log n)$                   & 1        & \cmark               \\\hline
            DBucket \cite{ngai_distributed_2024}                                                & $O(1)$                    & $O(1)$                    & $O(n)$                                     & $O(n\log n)$                   & 1        & \xmark               \\\hline

            \hline
            \multicolumn{7}{|c|}{\textbf{Other shufflers}}                                                                                                                                                                                                              \\\hline
            Rumor-based \cite{fanti_hiding_2017} ($\alpha=$ node degree)                        & $O(\alpha \cdot n)$       & $O(\alpha \cdot n)$       & $O(\alpha \cdot n)$                        & $O(\alpha \cdot n)$            & $\ge 1$  & \xmark               \\\hline
            NIDAR (2 layers) \cite{bunz_non-interactive_2021}                                   & $O(1)$                    & $O(1)$                    & $O(n)$                                     & $O(n^{1 + 1/2})$               & 1        & \xmark               \\\hline
            Secure-Agg.-based \cite{bell_secure_2020}                                           & $O(log^2n + n\log n)$     & $O(n \log^2 n)$           & $O(n^2 \log n)$                            & $O(n^2\cdot \log^2 n)$         & 1        & \xmark               \\\hline
        \end{tabular}
    \end{center}

    \textit{General notations}: $n =$ number of messages, $M =$ number of shuffling servers.

    \textit{Simplifications}: Message size is $O(1)$. Any security parameter $\lambda$ is $O(1)$. Setup costs are not included.
\end{table}

\begin{table}
    \footnotesize
    \begin{center}
        \renewcommand{\arraystretch}{1.2}
        \caption{Security comparison of the secure shuffling solutions}
        \label{tab:summary_security}
        \begin{tabular}{|p{0.17\linewidth}|c|p{0.25\linewidth}|p{0.05\linewidth}|p{0.05\linewidth}|p{0.05\linewidth}|p{0.2\linewidth}|}
            \hline
            \textit{Technique}                                                                  & $M$     & \multirow{2}{\linewidth}{\textit{Anonymous vs. \ldots}} & \multicolumn{2}{|p{0.10\linewidth}|}{\textit{Disruption resistance}} & \textit{Correct} & \textit{Other specificities}                                                                         \\
                                                                                                &         &                                                         & \textit{Sender}                                                      & \textit{Server}  &                              &
            \\\hline
            \multicolumn{7}{|c|}{\textbf{Mix networks} ($k$ = path length)}                                                                                                                                                                                                                                                                                          \\\hline
            Fixed-route \cite{chaum1981untraceable,danezis_mixminion_2003,freedman_tarzan_2002} & $\ge 1$ & $M-1$ malicious servers                                 & \cmark                                                               & \xmark           & \xmark                       &                                                                       \\\hline
            Free-route \cite{chaum1981untraceable,dingledine2004tor,chen_hornet_2015}           & $\gg 1$ & $k-1$ malicious servers per path                        & \cmark                                                               & \cmark           & \xmark                       & De-anonymization atk.                                                 \\\hline
            Advanced \cite{chaum_cmix_2017,piotrowska_loopix_2017}                              & $\gg 1$ & $k-1$ malicious servers per path                        & \cmark                                                               & \cmark           & \xmark                       &                                                                       \\\hline
            Trellis \cite{langowski_trellis_2023}                                               & $\gg 1$ & $f$\% of malicious servers                              & \cmark                                                               & \cmark           & \cmark                       & Trusted setup \& Weaker anonymity distribution                        \\\hline
            TORaaS   \cite{hohenberger_anonize_2014}                                            & 1       & $k-1$ malicious servers per path                        & \cmark                                                               & \cmark           & \xmark                       & De-anonymization atk.                                                 \\\hline

            \hline
            \multicolumn{7}{|c|}{\textbf{Verifiable shufflers}}                                                                                                                                                                                                                                                                                                      \\\hline
            Classic \cite{furukawa_efficient_2001,neff_verifiable_2001,bayer_efficient_2012}    & $>1$    & $M-1$ malicious servers                                 & \cmark                                                               & \cmark           & \cmark                       & Trusted setup                                                         \\\hline
            Riffle \cite{kwon_riffle_2016}                                                      & $>2$    & $M-2$ malicious servers                                 & \xmark                                                               & \xmark           & \cmark                       & Frequent setup w/ \cite{bayer_efficient_2012}                         \\\hline
            Atom \cite{kwon_atom_2017} ($k$: group size, $\alpha=\frac{k}{M}$)                  & $\gg 1$ & $k-1$ malicious serv. per group                         & \cmark                                                               & \cmark           & \cmark                       & Possibly a weaker anonymity distribution                              \\\hline
            Lattice-based \cite{aranha_verifiable_2023}                                         & $>1$    & $M-1$ malicious servers                                 & \cmark                                                               & \cmark           & \cmark                       & Trusted setup                                                         \\\hline

            \hline
            \multicolumn{7}{|c|}{\textbf{DC networks}}                                                                                                                                                                                                                                                                                                               \\\hline
            Dissent \cite{corrigan-gibbs_dissent_2010}                                          & $ >1$   & $M-1$ malicious servers                                 & \xmark                                                               & \xmark           & \xmark                       & Setup w/ \cite{bayer_efficient_2012}                                  \\\hline
            Riposte \cite{corrigan-gibbs_riposte_2015}                                          & $3$     & 1 malicious out of 3 servers                            & \cmark                                                               & \xmark           & \xmark                       &                                                                       \\\hline
            Spectrum \cite{newman_spectrum_2022}                                                & $2$     & $M-1$ malicious servers                                 & \cmark                                                               & \xmark           & \cmark                       & Setup uses another secure shuffler                                    \\\hline
            Sabre \cite{newman_spectrum_2022}                                                   & $3$     & $M-1$ malicious servers                                 & \cmark                                                               & \xmark           & \cmark                       &                                                                       \\\hline

            \hline
            \multicolumn{7}{|c|}{\textbf{MPC-based shufflers}}                                                                                                                                                                                                                                                                                                       \\\hline
            AsynchroMix \cite{lu_honeybadgermpc_2019}                                           & $\ge 4$ & $t < M/3$ malicious                                     & \cmark                                                               & \xmark           & \cmark                       &                                                                       \\\hline
            PowerMix \cite{lu_honeybadgermpc_2019}                                              & $\ge 4$ & $t < M/3$ malicious                                     & \cmark                                                               & \xmark           & \cmark                       &                                                                       \\\hline
            Blinder  \cite{abraham_blinder_2020}                                                & $\ge 5$ & $t < M/4$ malicious                                     & \cmark                                                               & \cmark           & \cmark                       &                                                                       \\\hline
            Clarion  \cite{eskandarian_clarion_2022}                                            & $\ge 3$ & $M-1$ malicious                                         & \cmark                                                               & \xmark           & \xmark                       &                                                                       \\\hline
            RPM  \cite{lu_rpm_2023}                                                             & $\ge 3$ & $t < M/2$ malicious                                     & \cmark                                                               & \cmark           & \cmark                       &                                                                       \\\hline
            Ruffle  \cite{a_ruffle_2023}                                                        & $= 3$   & $t < M/2$ malicious                                     & \cmark                                                               & \cmark           & \cmark                       &                                                                       \\\hline

            \hline
            \multicolumn{7}{|c|}{\textbf{TEE-based shufflers}}
            \\\hline
            Prochlo \cite{bittau_prochlo_2017}                                                  & $1$     & Malicious server                                        & \cmark                                                               & \xmark           & \cmark                       & Trusted hardware    \& \textbf{security issues} \cite{sasy_fast_2022} \\\hline
            ORShuffle \cite{sasy_fast_2022}                                                     & $1$     & Malicious server                                        & \cmark                                                               & \xmark           & \cmark                       & Trusted hardware                                                      \\\hline
            Butterfly Shuffle \cite{gu_efficient_2023}                                          & $1$     & Malicious server                                        & \cmark                                                               & \xmark           & \cmark                       & Trusted hardware                                                      \\\hline
            DBucket \cite{ngai_distributed_2024}                                                & $1$     & Malicious server                                        & \cmark                                                               & \xmark           & \cmark                       & Trusted hardware                                                      \\\hline

            \hline
            \multicolumn{7}{|c|}{\textbf{Other shufflers}}                                                                                                                                                                                                                                                                                                           \\\hline
            Rumor-based \cite{fanti_hiding_2017} ($\alpha=$ node degree)                        & $\ge 1$ & $t_\alpha$ malicious agents out of $N+M$                & \cmark                                                               & \cmark           & \cmark                       & De-anonymization attacks                                              \\\hline
            NIDAR (2 layers) \cite{bunz_non-interactive_2021}                                   & $1$     & Malicious server                                        & \xmark                                                               & \xmark           & \xmark                       & Trusted setup \& Weaker anonymity distribution                        \\\hline
            Sec.-Agg.-based \cite{bell_secure_2020}                                             & $1$     & Malicious server                                        & \xmark                                                               & \xmark           & \cmark                       &                                                                       \\\hline
        \end{tabular}
        \textit{General notations}: $n =$ number of messages, $M =$ number of shuffling servers.
    \end{center}
\end{table}

\paragraph{Complexity analysis}
To compare the scalability of the techniques, we systematically present communication and computation costs for both the servers and the data owners.
In Table \ref{tab:summary_efficiency}, the ``per-sender'' costs correspond to the (data owner) cost for one message sent.
The ``per-server'' costs correspond to the cost paid by each server to shuffle the whole batch of $n$ messages.
We emphasize the scalability in the number of messages $n$ and the number of servers $M$.

We make a few simplifying assumptions.
First, we assume the message size $L$ to be $O(1)$, but Subsection \ref{subsec:message-size} analyses the scalability with the message size.
Second, we assume that any security parameter $\lambda$ is $O(1)$; this parameter is present in cryptographic schemes to tune their security (i.e., larger implies more secure).
Third, the cost of the protocol setup is not part of the complexity analysis, because it is usually a one-time process.

\paragraph{Security analysis}

For each technique, we systematically identify which security properties are satisfied: anonymity, correctness, sender disruption resistance, and server disruption resistance.

As motivated in Section \ref{sec:def}, analysing a protocol anonymity is complex, so we focus three distinguishing criteria: threat model, anonymity distribution, and de-anonymization attacks.
Table \ref{tab:summary_security} has no dedicated column for anonymity distribution and de-anonymization attacks because these ``degrading factors'' only concern a few solutions.
A remark would appear in the column ``Other specificities'' if a technique has non-uniform distribution or known de-anonymization attacks.

Contrary to anonymity, correctness and disruption resistance are binary properties that are either present or absent (represented by check and cross marks in Table \ref{tab:summary_security}).

\subsection{Mix networks (mix nets)}
\label{subsec:comp_mix_net}
Mix networks are a well-known anonymization technique thanks to Tor \cite{dingledine2004tor} (a popular mix network used notably by activists and journalists for anonymous Web browsing).
The main idea of mix networks \cite{chaum1981untraceable} is to make a packet transit through several relay nodes before reaching its destination.
The packets are ``onion-encrypted'': one layer of encryption is added for each relay in the path.
Upon receiving the packet, each relay node removes one layer of encryption and forwards the decrypted payload to the next relay.
This mixing process prevents the relay nodes from linking the sender to the receiver.
For secure shuffling, all the messages are sent to the same agent (i.e., the analyzer or a public bulletin).

\begin{figure}
    \begin{subfigure}{0.45\linewidth}
        \centering
        \includegraphics[width=\linewidth]{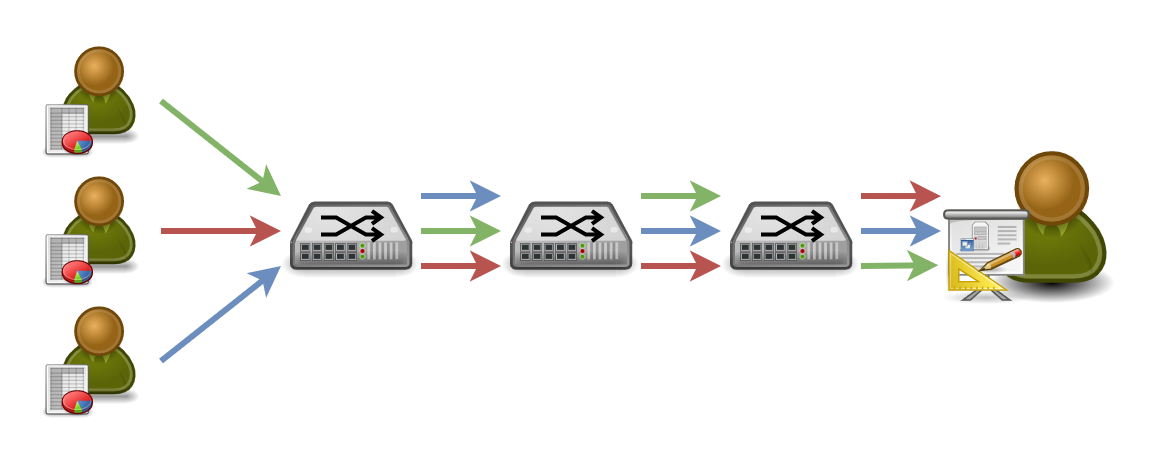}
        \caption{Fixed-route}
        \Description[Schema of a fixed-route mix net]{Schema of a fixed-route mix net: three data owners send their encrypted message to a chain of servers, the servers successively shuffles the ciphertexts and transmit the output to an analyzer.}
        \label{fig:fixed-route}
    \end{subfigure}
    \begin{subfigure}{0.45\linewidth}
        \centering
        \includegraphics[width=0.7\linewidth]{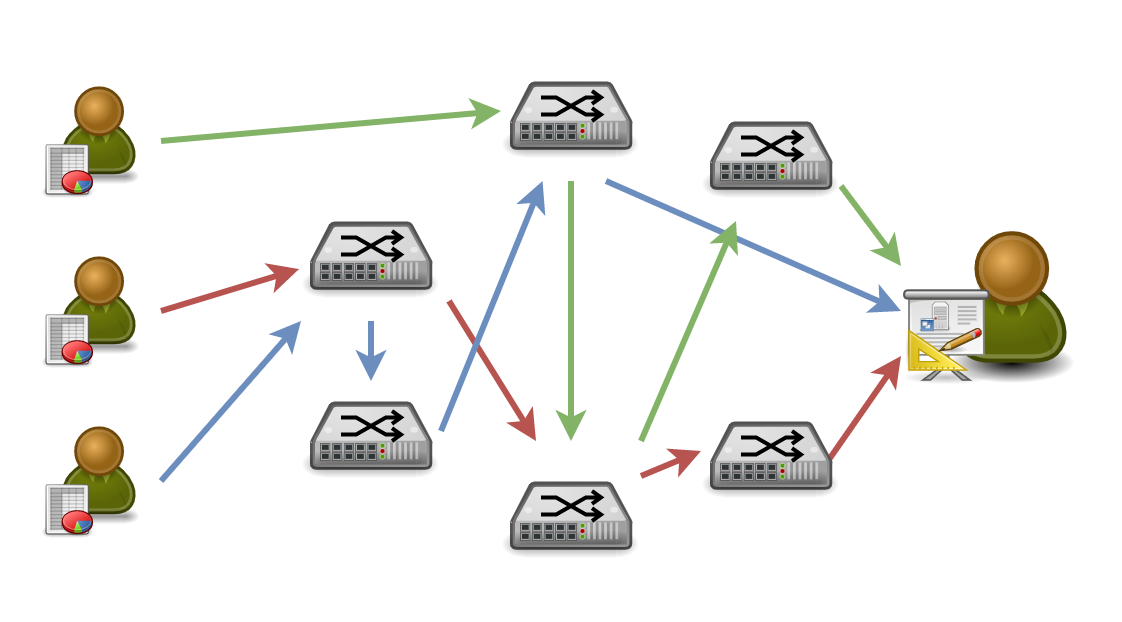}
        \caption{Free-route}
        \Description[Schema of a free-route mix net]{Schema of a free-route mix net: three data owners send their encrypted message through a network of server, each message takes a different pass and finishes on an analyzer.}
        \label{fig:free-route}
    \end{subfigure}
    \centering
    \caption{Mix networks}
\end{figure}

This anonymization technique is lightweight ($k$ public-key encryptions for a $k$-node path) and adds a small amount of latency due to the relays.
We distinguish free-route networks (Fig. \ref{fig:free-route}) where the sender freely chooses the path from fixed-route networks (Fig. \ref{fig:fixed-route}) where the path is pre-defined.
In the latter, the number of relays usually equals the number of servers.

\textbf{Remark about semi-trusted shuffler.}
Some papers, such as \cite{humphries_selective_2022}, use a semi-trusted shuffler that acts as a proxy.
A mix network being a proxy chain, the semi-trusted shuffler is a 1-node mix network with a non-colluding receiver.

We compare five types of mix networks: fixed-route, free-route, advanced, Trellis, and TORaaS.

\subsubsection{Scalability}
Mix networks are highly scalable because they require only lightweight encryption and decryption.
Some of them can even balance the load across the servers.
The sender cost is affordable even with resource-constrained devices.
Finally, the popularity of Tor is a concrete proof that it works at scale.

\subsubsection{Security}
Mix networks are anonymous as long as at least one server per relay path is honest.
They suffer from two types of de-anonymization attacks based on their design: traffic analysis attacks (i.e., a global adversary observing to trace packets path) and replay attacks (i.e., a node forwarding twice the same packet and observing which message appears twice in the shuffle output).
While some mix nets resist traffic analysis attacks, replay attacks are a general problem that can be mitigated using an anonymous authentication system.

Mix networks (except Trellis) guarantee no correctness because any relay node can drop packets (and then censor some of the senders).
However, they have sender disruption resistance.

\subsubsection{Noticeable solution: Tor as a Shuffler (TORaaS)}
Let us analyze a deployed secure shuffler that was never formally introduced in literature but implicitly present in some papers: TORaaS (i.e., shuffling through the Tor anonymization network).
In TORaaS, we use the existing Tor network instead of deploying a new mix network dedicated to the shuffling task.
The shuffler traffic is mixed with all Tor traffic.
The practitioner only needs to deploy a Web server accessible via Tor.

We distinguished this solution from other free-route mix networks because Tor has some specific and well-documented flaws which are important to discuss.
This solution is so simple and attractive that it must be analyzed independently so a reader can choose it safely.

While this solution is attractive due to its simplicity, practitioners should carefully consider the possibility of de-anonymization attacks \cite{arp_torben_2015}.
Even though there exists mitigation techniques, Tor is still particularly sensitive to such attacks as it is a free-route mix network.

\subsection{Dining Cryptographers networks (DC nets)}
These solutions take inspiration from the Dining Cryptographers problem \cite{chaum_dining_1988,golle_dining_2004}.
This anonymity problem is named after a scenario where three cryptographers are dining together and learn that their meal has been paid for, but they want to know whether one of them paid or if it was the NSA; without revealing who paid if it was one of them.

\begin{figure}
    \centering
    \includegraphics[width=.3\linewidth]{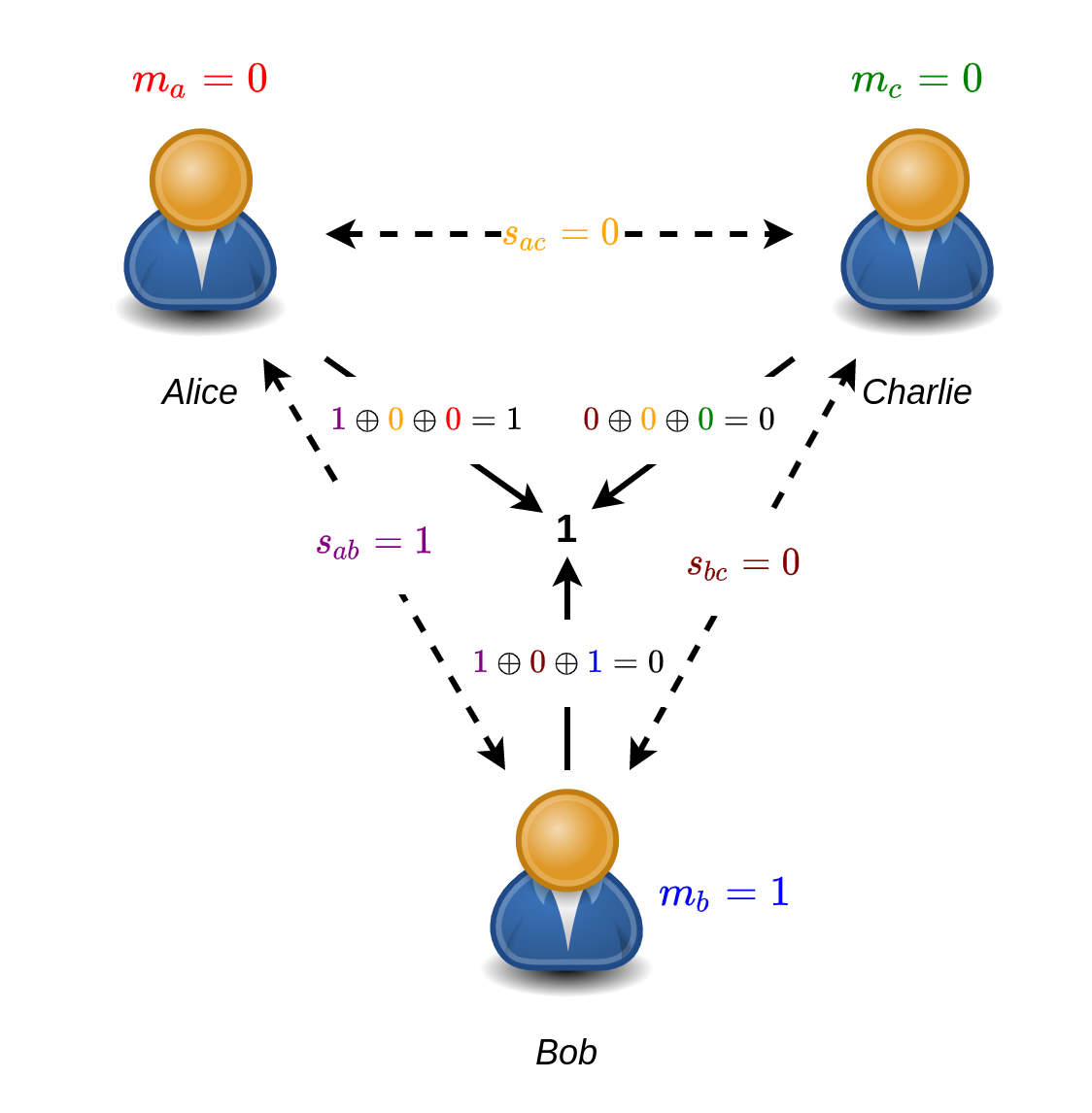}
    \caption{Dining Cryptographers problem}
    \Description[Schema of the DC problem]{Schema of the DC problem: three secret owners exchange additive masks with each other and then publish their mask values to reveal the final result.}
    \label{fig:dc-prob}
\end{figure}

The solution proposed by \citet{chaum_dining_1988} is to exchange complementary binary masks between the participants (as represented in Fig. \ref{fig:dc-prob}).
Then, each participant publishes a masked value.
Once aggregated, the masked values reveal the anonymous bit of information.
This masking trick was reused as a foundation for several anonymous protocols referred to as ``DC nets''.
Most anonymization systems based on DC nets complement DC nets with other cryptographic solutions to combine scalability and security.
Recently, Shirali et al. \cite{shirali_survey_2022} wrote a survey focused on DC nets.

We identified four major DC nets: Dissent \cite{wolinsky_dissent_2012}, Riposte \cite{corrigan-gibbs_riposte_2015}, Spectrum \cite{newman_spectrum_2022}, and Sabre \cite{vadapalli_sabre_2022}.

\subsubsection{Scalability}
Until 2012 \cite{wolinsky_dissent_2012}, DC nets had been considered a purely theoretical construction compared to the highly scalable mix networks.
Dissent \cite{wolinsky_dissent_2012} marked a turning point because practical applications could be imagined, but scalability was still very limited.
Recent solutions, such as Riposte \cite{corrigan-gibbs_riposte_2015} and Spectrum \cite{newman_spectrum_2022}, are the first scalable DC nets.
Riposte was the first anonymous broadcaster able to shuffle 1M messages within a few hours.
It is not as scalable as mix networks but requires fewer servers (i.e., 2 or 3).

\subsubsection{Security}
DC nets were initially proposed to offer theoretically proven anonymity.
Dissent is the only surveyed technique hiding the size and number of sent messages by each user.
This leakage avoidance (and its cost) is discussed in Subsection \ref{subsec:ultimate_anon}.

Each DC net has a rather unique architecture, so they do not share many common points, except their foundation: the DC problem.
For simplicity, we only highlight the properties of the state-of-the-art method: Spectrum \cite{newman_spectrum_2022}.
It is most optimized for two servers and is anonymous against one malicious server.
There are no known de-anonymization attacks specific to Spectrum (or DC nets in general).
Spectrum has correctness and is resistant to sender disruption only.

\subsection{Verifiable shufflers}
Verifiable shufflers are also called verifiable mix nets because they can be represented as fixed-route mix networks (see Fig. \ref{fig:fixed-route}) where each node must prove it performed a shuffle.
Verifiable shufflers are a line of work distinct from mix networks because they focus on the shuffle proof more than on the network architecture.
For example, the encryption schemes and the applications are very different.
Verifiable shufflers are usually small networks of mix nodes performing expensive computations (i.e., proofs and verifications).

For each shuffle, a mix node produces a Zero-Knowledge Proof (ZKP), proving the correctness of the shuffle.
Before the shuffle, the mix node commits to a secret random permutation and the proof shows that this random permutation was respected.
Verifiable shufflers batch the incoming messages before applying the permutation.
This batching protects against traffic analysis attacks.

For more details about verifiable shufflers, we recommend the survey paper of \cite{haines_sok_2020}.

\begin{figure}
    \centering
    \includegraphics[width=.4\linewidth]{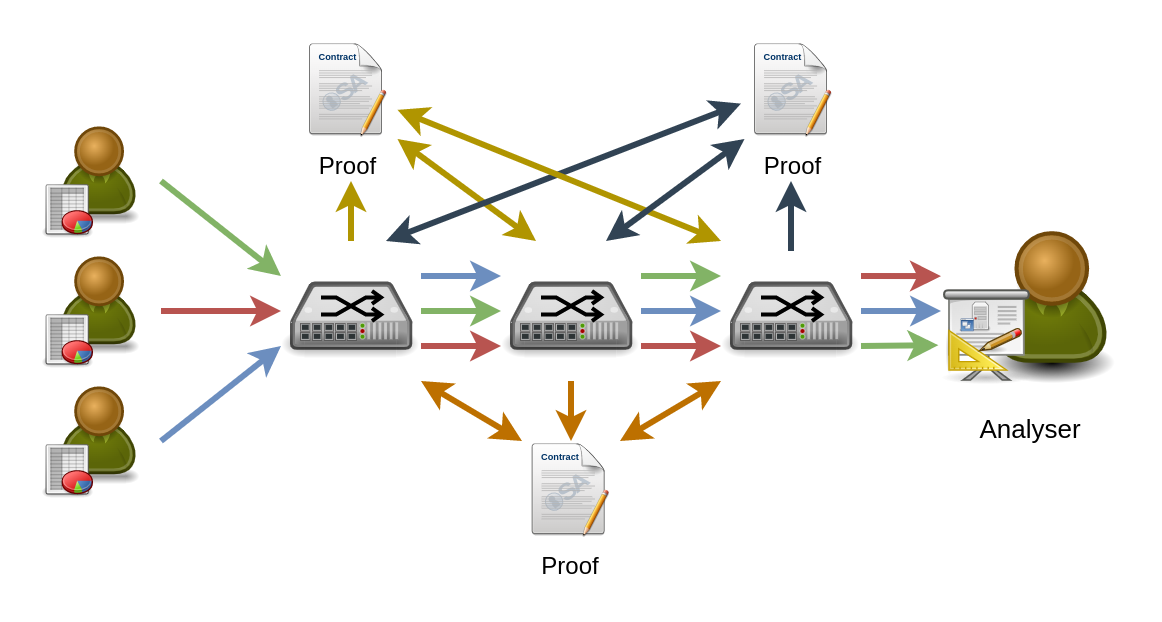}
    \caption{Classic verifiable shufflers}
    \Description[Schema of a verifiable shuffler]{Schema of a verifiable shuffler: three data owners send their encrypted message to a chain of servers, the servers successively shuffles the ciphertexts, prove the shuffle and verify each other proofs. Finally, they transmit the final output to an analyzer.}
    \label{fig:verif_shuffler}
\end{figure}

We identify four major verifiable shufflers in the literature: Classic verifiable shufflers \cite{bayer_efficient_2012}, Riffle \cite{kwon_riffle_2016}, Atom \cite{kwon_atom_2017}, lattice-based verifiable shuffle \cite{aranha_verifiable_2023}.

We use \citet{bayer_efficient_2012} as reference for classic verifiable shuffling, but some recent works \cite{hebant_linearly-homomorphic_2020,fauzi_efficient_2017} have described various optimizations such as homomorphic proofs \cite{hebant_linearly-homomorphic_2020} (making the verification cost constant in the number of nodes in the network).

Recently, \citet{balle_amplification_2023} used verifiable shufflers to build an ``alternating shuffler.''
Their protocol splits the messages to shuffle into several subsets and shuffle them independently.
An alternating shuffler outputs a permutation that is imperfect, but good enough to provide DP guarantees.

\subsubsection{Scalability}
The classic verifiable shufflers \cite{bayer_efficient_2012} have some scalability bottlenecks (i.e., cannot scale above tens of thousands of messages), but it does not prevent concrete use cases.
However, the most recent adaptations of verifiable shufflers to anonymous communications, namely Riffle \cite{kwon_riffle_2016} and Atom \cite{kwon_atom_2017}, do not have this scalability issue.
For example, Atom can shuffle a million message in less than an hour.

\subsubsection{Security}
They are anonymous as long as one of the servers is honest.
Thanks to the ZKP, correctness is guaranteed contrary to mix networks.
Only Riffle is not resistant to sender disruption, while Atom is the only verifiable shuffler with server disruption resistance.
No de-anonymization attack specific to verifiable shufflers is known.

\subsection{MPC-based shufflers}

An MPC-based shuffler is a group of $M$ servers that securely collaborate to output shuffled messages.
The data owners, willing to anonymously broadcast a message, send a share of their secret message to each server.
The servers then use an MPC protocol to obtain the shuffled messages.

\begin{figure}
    \centering
    \includegraphics[width=.4\linewidth]{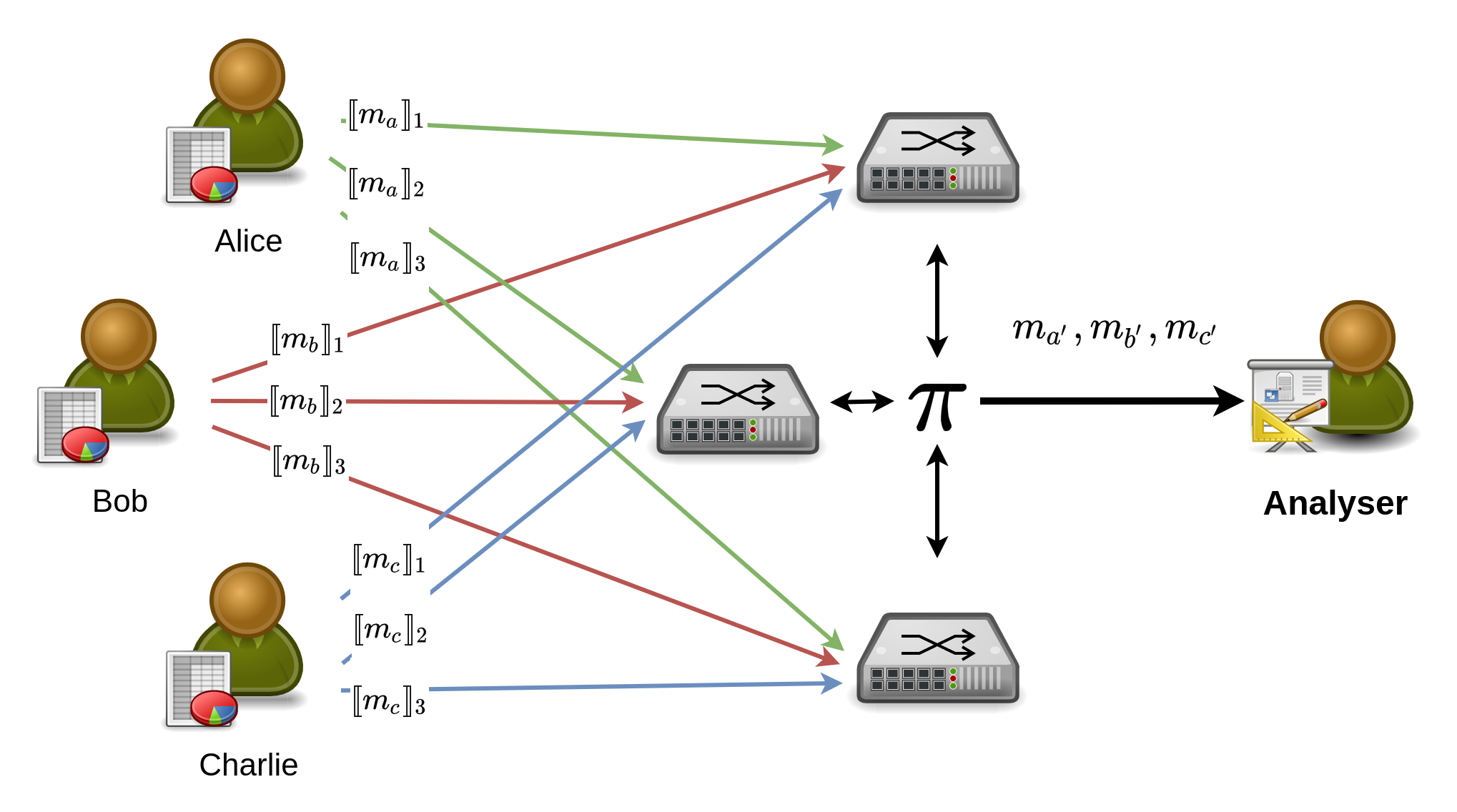}
    \caption{MPC-based secure shufflers ($\pi$: an MPC protocol)}
    \label{fig:mpc-shufflers}
    \Description[Schema of an MPC-based shuffler]{Schema of an MPC-based shuffler: three data owner send one share of their secret to three different servers. The servers then perform an MPC shuffle program and transmit the output to an analyzer.}
\end{figure}

We identified six major MPC-based shufflers in the literature: AsynchroMix \cite{lu_honeybadgermpc_2019}, PowerMix \cite{lu_honeybadgermpc_2019}, Blinder \cite{abraham_blinder_2020}, Clarion \cite{eskandarian_clarion_2022}, RPM \cite{lu_rpm_2023}, and Ruffle \cite{a_ruffle_2023}.

Some of these shufflers (e.g., Clarion) are built upon generic ``secret-shared shuffle protocols.''
This generic building block could be further optimized thanks to recent advances \cite{chase_secret-shared_2020,gao_multiparty_2024,song_secret-shared_2024}.

\subsubsection{Scalability}
PowerMix and AsynchroMix hit their scaling limits at a few thousand messages.
However, the most recent solutions (Blinder, RPM, Clarion, and Ruffle) can scale up to millions of shuffled messages (at a time).
Blinder even leverages GPUs to speed up the process.

\subsubsection{Security}
All these techniques (except Clarion) share similar security properties: anonymous if there is a majority of honest servers, correctness, and sender disruption resistance.
Blinder even has server disruption resistance.
On the other hand, Clarion only has sender disruption resistance (i.e., no correctness and no server disruption resistance), but it is anonymous against $M-1$ malicious servers.
No de-anonymization attack specific to these solutions has been yet described.

\subsection{TEE-based shufflers}
Bittau et al. \cite{bittau_prochlo_2017} introduced the first TEE-based secure shuffler.
The only step to conceive a TEE-based secure shuffler is to provide a shuffle code executable on the chosen architecture (e.g., Intel SGX \cite{intel_software_2015}).
Shuffling requires simple operations, but the implementation must be oblivious: the server owner should not be able to infer information about the data based on the memory access patterns.
In other words, TEE-based shufflers behave as a proxy in a secure environment that guarantees data confidentiality.
We identified four TEE-based secure shufflers: Prochlo \cite{bittau_prochlo_2017}, ORShuffle \cite{sasy_fast_2022}, the Butterfly Oblivious Shuffle \cite{gu_efficient_2023}, and DBucket \cite{ngai_distributed_2024}.

\begin{figure}
    \centering
    \includegraphics[width=.35\linewidth]{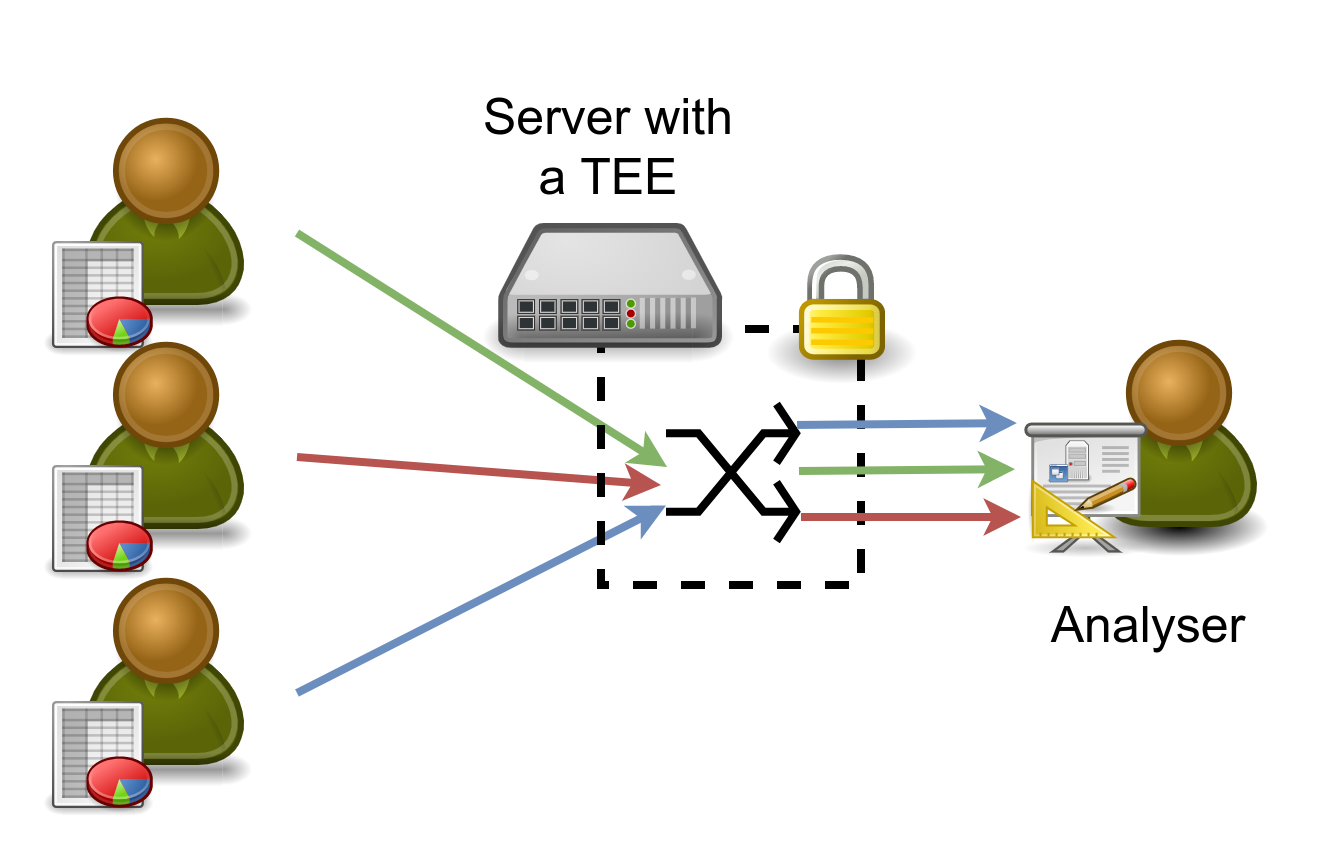}
    \caption{TEE-based secure shufflers}
    \label{fig:tee-shufflers}
    \Description[Schema of a TEE-based shuffler]{Schema of a TEE-based shuffler: three data owner send their encrypted message to the secure enclave of a server, a shuffle is performed in the secure enclave and the result is transmitted to an analyzer.}
\end{figure}

\subsubsection{Scalability}
Prochlo uses a ``stash shuffle'' with an asymptotic cost equivalent to a trusted shuffler (e.g., $O(n)$ computations and communications for the server, and $O(1)$ for the sender).
ORShuffle achieves a server cost of $O(n\log^2n)$, and the Butterfly Shuffle and DBucket $O(n\log n)$.

\subsubsection{Security}
TEE is satisfying as long as the user fully understands the implications of the trust assumption made.
Subsection \ref{subsec:tee-strong-assumptions} discusses the strength of this assumption.

Moreover, flawed TEE implementations can be sensitive to side-channel attacks.
For example, Sasy et al. \cite{sasy_fast_2022} highlighted that Prochlo is sensitive to some recent attacks.

In the absence of side-channel attacks, TEE-based shufflers are anonymous against a malicious server.
They satisfy shuffle correctness and sender disruption resistance.
However, there is no server disruption resistance since the server owner can disconnect the server to disrupt the shuffle.

\subsection{Other shufflers}
Finally, we identified three other secure shufflers that do not fit in any of the previous categories: rumor-based shuffler \cite{fanti_hiding_2017}, NIDAR (Non-Interactive Differentially Anonymous Router) \cite{bunz_non-interactive_2021}, and secure-aggregation-based shuffler \cite{bell_secure_2020}.

Both rumor-based and secure-aggregation-based solutions are present in this survey for completeness.
They are not practical and offer no scalability or security improvements compared to the rest of the secure shufflers surveyed in our paper.

NIDAR is particularly interesting because it is a single server solution (based on a cryptographic primitive called ``Functional Encryption'') that does not require a secure enclave.
Its scalability and security are imperfect, but it opens this direction with promising first results, especially its DP-inspired relaxation of anonymity definition.
The main weaknesses of NIDAR are the lack of sender disruption resistance and the need for a trusted setup.

\subsection{Scalability with the message size}
\label{subsec:message-size}
This last comparison subsection studies the scalability of the techniques with the message size because message sizes can vary significantly in privacy-preserving computations: from small integers to large vectors of floating-point numbers.
Most techniques have a linear dependency on the message size, but a few have different relationships.

The comparison is summarized in Table \ref{tab:summary_msg_size}.
We denote the original cost of a given technique as $\mathcal{C}$.
Hence, a cost of $\mathcal{C}\cdot O(L)$ means all the original costs are multiplied by $O(L)$.

The computation costs of proof and verification protocols usually correspond to the number of modular exponentiations needed.
Since modular exponentiations have a cubic cost, it induces a $O(L^3)$ dependency in classic verifiable shufflers \cite{bayer_efficient_2012} and Atom \cite{kwon_atom_2017}.

All the DC-nets have a linear dependency on the message size.
For Riposte, Spectrum, and Sabre, an additive term (linear in the message size) appears because of a cryptographic primitive they use (i.e., Function Secret Sharing \cite{gilboa_distributed_2014}).
All the TEE-based and MPC-based shufflers have a linear dependency on the message size.
Among the other shufflers, only secure-aggregation-based shufflers have a non-linear dependency on message size.

\begin{table}
    \footnotesize
    \begin{center}
        \renewcommand{\arraystretch}{1.5}
        \caption{Scaling of the surveyed techniques with the message size}
        \label{tab:summary_msg_size}
        \begin{tabular}{|p{0.25\linewidth}|p{0.11\linewidth}|p{0.15\linewidth}|p{0.11\linewidth}|p{0.15\linewidth}|}
            \hline
            \textit{Technique}              & \multicolumn{2}{|c|}{\textit{Per-Sender}}                 & \multicolumn{2}{|c|}{\textit{Per-Server}}                                                           \\
                                            & \textit{Comm.}                                            & \textit{Computations}                     & \textit{Comm.}          & \textit{Computations}         \\ \hline
            Classic. verif. shufflers, Atom & $\mathcal{C}\cdot O(L)$                                   & $\mathcal{C}\cdot O(L^3)$                 & $\mathcal{C}\cdot O(L)$ & $\mathcal{C}\cdot O(L^3)$     \\\hline
            Riposte, Spectrum, Sabre        & $\mathcal{C} + O(ML)$                                     & $\mathcal{C} + O(L)$                      & $\mathcal{C} + O(nL)$   & $\mathcal{C}\cdot O(L)$       \\\hline
            Sec.-Agg.-based                 & $\mathcal{C} + O(nL)$                                     & $\mathcal{C} + O(nL\log n)$               & $\mathcal{C} + O(n^2L)$ & $\mathcal{C} + O(n^2L\log n)$ \\\hline \hline
            \textbf{Rest of the shufflers}  & \multicolumn{4}{|c|}{\normalsize $\mathcal{C}\cdot O(L)$}                                                                                                       \\\hline
        \end{tabular}
    \end{center}

    \textit{Notations}: $L$ is the message size, $\mathcal{C}$ is the ``original cost'' of the technique (see Table \ref{tab:summary_efficiency}).
\end{table}

%% file: sec/applications.tex
\section{Applications of secure shufflers to privacy-preserving computations}
\label{sec:applications}

\subsection{Private statistics}
Following the first anonymous communication systems, several works \cite{brickell_efficient_2006,durak_precio_2023,gunther_pem_2020,hohenberger_anonize_2014} have built systems based on message anonymization to compute private aggregate statistics (e.g., mean).
As detailed in Section \ref{sec:def}, the sender anonymization layer used in these work can be seen as a shuffler.
This equivalence between message anonymization and shuffling is common in Differential Privacy where papers alternatively use ``anonymous messages'' \cite{ghazi_power_2021,ghazi_private_2020} and ``shuffled messages'' \cite{girgis_shuffled_2020,erlingsson_encode_2020}.

On the one hand, some systems \cite{brickell_efficient_2006,hohenberger_anonize_2014} relied on shuffler to anonymize the message sender.
For example, Brickell et al. \cite{brickell_efficient_2006} built a privacy-preserving data collection system based on such message anonymization.
In their system, each data owner sends their private data to the shuffler and the shuffler outputs a set of ``anonymous'' messages.
This approach provides a weak privacy preservation as it does not mitigate any privacy leakage caused by the message content.

On the other hand, other works \cite{durak_precio_2023,gunther_pem_2020} built inspired by Ishai et al. \cite{ishai_cryptography_2006}: they require each data owner to generate secret shares and to send these shares to the shuffler.
As explained in Section \ref{subsec:ikos}, the secret sharing will prevent the secret reconstruction.
Thus, these works provided stronger privacy guarantees than \cite{brickell_efficient_2006,hohenberger_anonize_2014} because they mitigate any privacy leakage from individual messages; only the aggregated result is revealed.

\subsection{Differentially Private computations}
\label{subsec:shuffle_dp}
While private statistics systems such as \cite{durak_precio_2023,gunther_pem_2020} already provide strong privacy guarantees, works in DP proposes an even stronger privacy amplification mitigating privacy leakage both from the shuffled messages and from the aggregated output.

\begin{figure}
  \centering
  \includegraphics[width=0.6\linewidth]{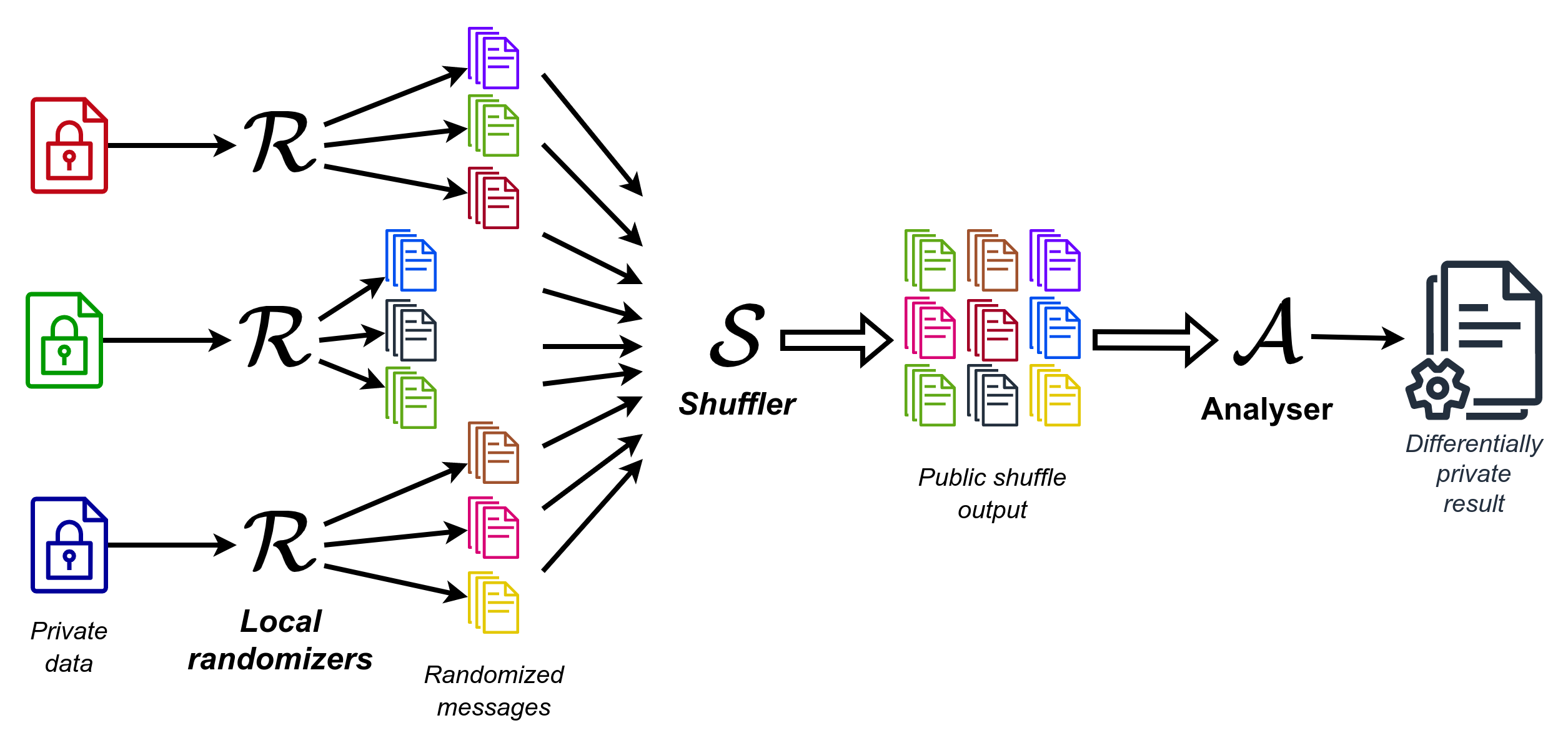}
  \caption{Shuffle Differential Privacy: (1) The data owners split their private value into several random messages using the randomizer $\mathcal{R}$, (2) they transmit these messages to $\mathcal{S}$, (3) the shuffler outputs a random permutation of the messages and (4) the analyser can perform any relevant operation on these messages.}
  \label{fig:shuffle_dp}
  \Description[Schema of the Shuffle DP framework]{Schema of the Shuffle DP framework: three private data samples are each split into 3 messages by a local randomizer. The messages are transmitted to a shuffler that shuffles them and transmit the shuffle output to an analyzer.}
\end{figure}

This DP framework based on shuffler is called ``the shuffle DP model'' \cite{ghazi_private_2020-1} and consists of two randomized algorithms (see Fig. \ref{fig:shuffle_dp}):
\begin{itemize}
  \item A local randomizer $\mathcal{R}: \mathcal{X} \rightarrow \mathcal{Y}$
  \item A shuffler $\mathcal{S}: \mathcal{Y}^*  \rightarrow \mathcal{Y}^*$ that randomly permute its inputs.
\end{itemize}

The protocol $\mathcal{P}=\mathcal{S}\circ\mathcal{R}$ is $\pt{\epsilon, \delta}$-shuffle differentially private if $\mathcal{S}\pt{\mathcal{R}(x_1)\dots\mathcal{R}(x_n)}$ is $\pt{\epsilon, \delta}$-differentially private.
The output  $\mathcal{S}\pt{\mathcal{R}(x_1)\dots\mathcal{R}(x_n)}$ can then be used by any analyzer  $\mathcal{A}: \mathcal{Y}^* \rightarrow \mathcal{Z}$ to provide a differentially private result (e.g., a histogram or a sum).

Several papers \cite{balle_privacy_2019,chen_generalized_2024,erlingsson_amplification_2019,feldman_stronger_2023} showed that the shuffling induces a privacy amplification because it requires less noise than local DP to offer comparable privacy guarantees.
The privacy-accuracy trade-off of Shuffle DP approaches the trade-off of Central DP without requiring a trusted party.
This good trade-off without a trusted party assumption often motivates the use of shuffle DP in privacy-preserving computations.
For an extensive overview on the benefits of Shuffle DP compared to local and central DP, we refer to Cheu \cite{cheu_differential_2021}.

While shuffle DP is usable for any statistical computation, it provides a privacy-utility close to central DP mostly for sum operations.
The first shuffle DP papers focused on private summation \cite{balle_improved_2019, balle_private_2020,ghazi_pure_2020}, but recent works described more complex statistical operations: private quantiles \cite{aamand_lightweight_2025}, private uniformity testing \cite{balcer_connecting_2021,cheu_differential_2021}, private histograms \cite{balcer_separating_2020,bell_distributed_2022,wang_beyond_2025} and other related computations \cite{ghazi_power_2021,luo_frequency_2022,humphries_selective_2022}.

\subsection{Federated Learning and Secure Aggregation}
Federated Learning (FL) \cite{kairouz_advances_2021} is a decentralized ML paradigm enabling to collaboratively train an ML model without exchanging their (private) training data.
While there exists several FL variants, the most popular setup is ``horizontal FL:'' each data owner owns a small dataset over the same set of features.

\begin{figure}
  \centering
  \includegraphics[width=0.45\linewidth]{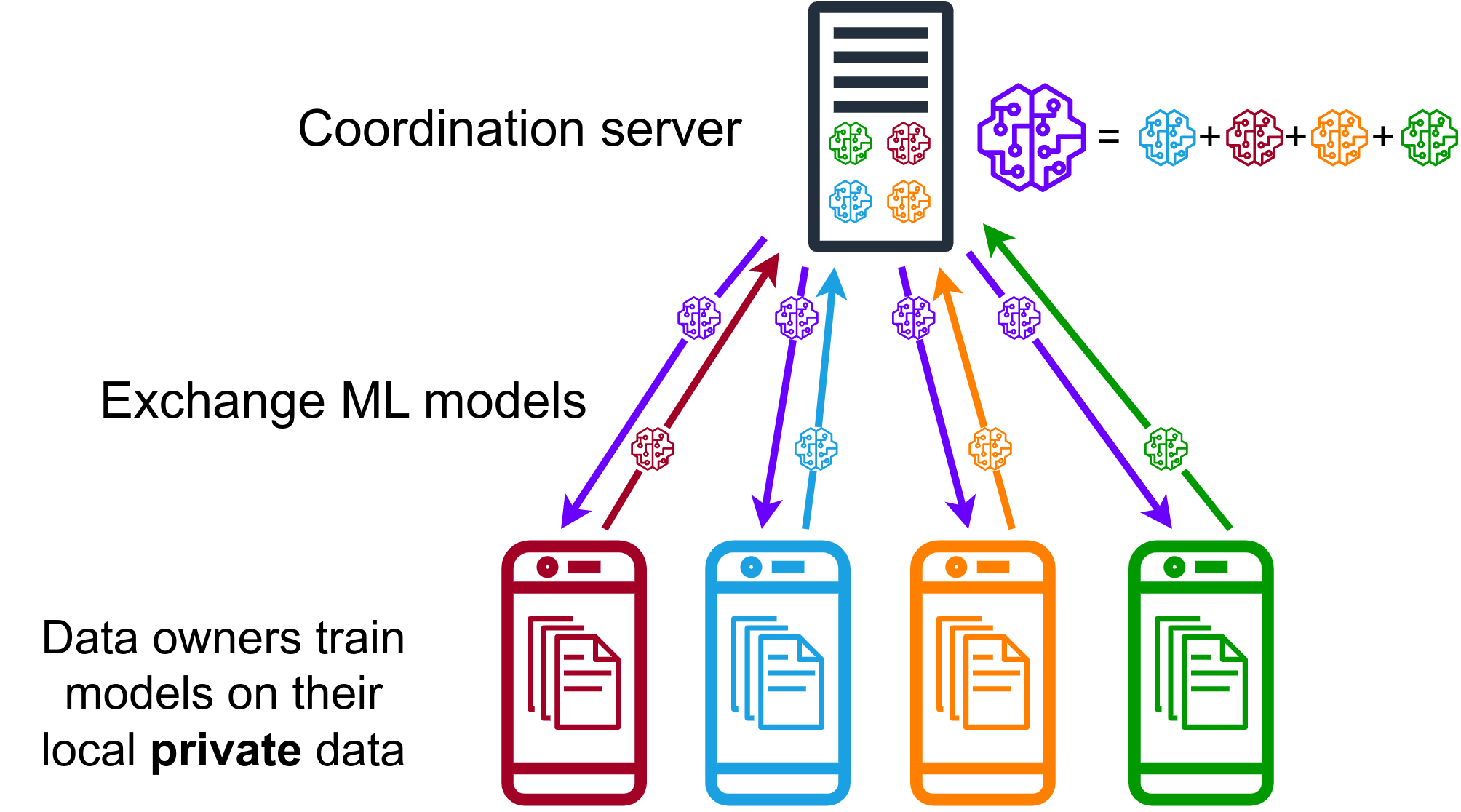}
  \caption{High-level Federated Learning architecture}
  \Description[Schema of an FL architecture]{Schema of an FL architecture: multiple data owners each trains locally an ML model and then communicate with a group of coordination servers outputting a global ML model}
  \label{fig:FL_arch}
\end{figure}

Traditional FL protocols are iterative decentralized algorithms with each iteration having the following structure: (1) Each data owner trains a local ML model on its private dataset, (2) Each data owner sends its local model to a server, (3) the server aggregates the local model and sends back a global model.
Fig. \ref{fig:FL_arch} sketches a high-level representation of FL.

Such aggregation-based FL protocols enable to train a wide variety of ML models: any model which training relies on gradient descent-based \cite{konecny_federated_2016} (e.g., neural networks or linear regression), k-means protocols \cite{zhang_practical_2022,li_differentially_2023,holzer_dynamically_2023,kumar2020federated}, Singular Vector Decomposition \cite{chai_federated_2021}, random forests \cite{de_souza_dfedforest_2020,liu_federated_2020}, and Gradient Boosted Decision Trees \cite{tian_federboost_2020,cheng_secureboost_2021}.

Even though FL does not require private data exchanges, the ML models (or gradients) exchanged during these protocols can leak private information \cite{nasr_comprehensive_2019,dartel_evaluating_2025}.
To enhance privacy, many works \cite{kairouz_advances_2021} have relied on secure aggregation to obtain the sum of the local models without revealing the individual models.

\paragraph{Secure aggregation}
Secure aggregation is a protocol taking as input $n$ private vectors $v_i$ and outputting $\sum_{i=1}^n v_i$ without revealing the input vectors.
Several solutions have been proposed to implement secure aggregation including \emph{additive masking} \cite{bonawitz_practical_2017,bell_secure_2020,sabater_accurate_2021,bell_acorn_2022}, \emph{Homomorphic Encryption} \cite{shi_privacy-preserving_2011}, \emph{Trusted-Execution Environment} (TEE) \cite{lie_glimmers_2017}, and \textbf{shuffling-based} \cite{ishai_cryptography_2006}.
\citet{mansouri_sok_2023} surveyed secure aggregation protocols.

We can implement \textbf{shuffling-based secure aggregation} in two manners.
On the one hand, we can perform an ``exact'' secure aggregation using the protocol proposed by \citet{ishai_cryptography_2006} (see in Section \ref{subsec:ikos}).
On the other hand, we can perform a ``differentially private'' secure aggregation using the DP summation mentioned in Section \ref{subsec:shuffle_dp}.
Several FL papers have recently used this shuffle-DP aggregation \cite{girgis_shuffled_2021,liu_flame_2021,xu_camel_2024}.

\paragraph{Secure sparse aggregation}
There is a growing interest in gradient sparsification to reduce communications in FL \cite{stich_sparsified_2018}.
As for classic secure aggregation, the sparse secure aggregation can replace the non-private aggregation step to enhance privacy.
Luo et al. \cite{luo_frequency_2022} presented a \textbf{shuffling-based sparse vector aggregation}.

\subsection{Towards new shuffling-based applications}
Previous applications all use secure shufflers to perform a form of private data aggregation.
However, secure shufflers can also find applications in other types of privacy-preserving protocol.

For example, \citet{wang_beyond_2025} recently proposed a shuffling-based taxi-hailing service.
In this application, users willing a taxi can submit their request and location through a secure shuffler and the secure shuffler will provide DP guarantees on their location data.
More generally, the authors introduced the notion of ``private individual computations:'' a broad range of ``permutation-equivariant'' computations such as taxi-hailing and information retrieval.
In such protocols, each user receives a personalized service/output, while secure shufflers amplify the privacy.
Thus, their work opened the way to new applications of secure shufflers.

Another recent example is the shuffling-based private information retrieval described by \cite{gascon_computationally_2024}.

\subsection{Applications unrelated to privacy-preserving computations}
Our survey focuses specifically on application to privacy-preserving computations as they are the protocols that revived the interest in secure shufflers.
However, there exist other ``historical'' applications that we do not survey: web anonymization \cite{dingledine2004tor}, anonymous messaging \cite{sasy_sok_2023}, electronic voting \cite{neff_verifiable_2001,juels_coercion-resistant_2010}, and blockchain operations \cite{kutylowski_coinshuffle_2014}.

%% file: sec/how_to_choose.tex
\section{How to choose a secure shuffler?}
\label{sec:how_to_choose}
Our survey presents a wide range of possible shufflers, but there is no absolute best choice.
The goal of this section is to guide ML/DP researchers in choosing a secure shuffler for each use case.
We first present the guidelines and then apply to three concrete use cases.

\textbf{Scalability trilemma}: shuffle time, message size, and number of messages.
In theory, any shuffler can shuffle numerous large messages, but it may take ages.
In these guidelines, we simplify this time-size-number trade-off and focus on the number and size of the messages.
Implicitly, we consider a secure shuffler suitable if it can shuffle the messages within a few hours.
This assumption is in line with the security literature, which rarely provides experimental results of shuffles exceeding 12 hours, such long shuffle times are usually associated with unsuccessful methods.

\subsection{Guidelines}
As guidelines, we propose a list of eight simple questions.
These guidelines are more a ``rule of thumb'' than a formal method, but they should help quickly identify one or several suitable solutions for a use case.
The practitioner must refer to Tables \ref{tab:summary_efficiency}, \ref{tab:summary_security}, and \ref{tab:summary_msg_size} to filter the shufflers based on their answers.
This subsection presents the guideline questions one by one, explains their meaning and the kind of expected answer.

\subsubsection{Question 1: How many messages are sent during one shuffling round?}
Possible answers: Less than 1K, Less than 100K, Less than 1M, More than 1M.

\emph{Meaning:}
Two factors influence the number of messages: the number of data owners, and the private computation task.
First, the influence of the number of data owners is straightforward.
Second, the type of private computation tasks may produce more or less private messages to shuffle: in gradient aggregation for FL, each message contains a whole gradient, while in histogram-making, each item might be shuffled independently

\paragraph{Rule of thumb}
Since the asymptotic costs reported in Table \ref{tab:summary_efficiency} are hard to relate to a concrete number of messages, we propose a \emph{rule of thumb} to link the asymptotic costs to a number of shuffled messages within a few hours.
We propose four thresholds to categorize the techniques based on their scalability with the number of messages.
These thresholds are inspired by the observations of the experimental results reported in the surveyed papers.

\emph{With less than 1K messages}, we can use any technique.
\emph{With less than 100K messages} (but more than 1K), we must avoid all techniques with a server cost above $O(n^2)$.
For example, PowerMix must be avoided (AsynchroMix being tightly related to it must also be avoided).
\emph{With less than 1M messages} (but more than 100K), we must avoid all techniques with a server cost above $O(n\log n)$.
For example, Spectrum is no longer suitable.
Note that Blinder is an exception that can leverage GPU computations to scale to millions of messages.
\emph{For more than 1M messages}, we recommend using only linear (or sublinear) techniques in $n$.
For example, Trellis scales to dozens of millions using load balancing.

\subsubsection{Question 2: How large are the messages?}
Possible answers: Less than 1KB, few KBs, few MBs, More than 10 MBs.

\emph{Meaning:}
The main factor influencing the message size is the private computation task because it defines the content of the messages.
For example, in gradient aggregation, gradients can produce large messages, while histogram-making requires small messages if we shuffle the items independently (i.e., one item = one message).

\paragraph{Rule of thumb}
Similarly to the previous guideline question, we propose a rule of thumb to simplify the interpretation of the asymptotic costs.
We propose four thresholds to compare the scalability with the message size (refer to Table \ref{tab:summary_msg_size} complexity comparison).
\emph{With less than 1KB per message}, any technique should be suitable.
\emph{With few KBs per message}, we should avoid techniques that are polynomial in the message size (e.g., classic verifiable shufflers).
\emph{With few MBs per message}, we should use techniques scaling linearly with the message size and with a sub-quadratic scaling with the number of messages.
For example, this rule filters out PowerMix.
\emph{For more than a few MBs per message}, one will need to look among the very few techniques compatible with activities such as anonymous whistleblowing requiring large message exchanges (e.g., Riffle).

\subsubsection{Questions 3: Are the data owners resource-constrained?}
Possible answers: Very constrained, Slightly constrained, or no.

\emph{Meaning:}
If the answer to this question is ``very constrained'' (e.g., for smartphones), one should avoid all solutions with a sender cost dependent on $n$.
For example, Blinder and Spectrum would be filtered out due to their sender costs.
If the answer is ``slightly constrained'', one should avoid all solutions with a sender cost with a dependency more than logarithmic on $n$.
For example, Blinder has a square-root dependency and would be filtered out, while Spectrum has a cost logarithmic in $n$, which is acceptable in this case.

\subsubsection{Questions 4: Is there a trusted third party?}
Possible answers: Yes or No.

\emph{Meaning:}
In some protocols, a trusted third party (e.g., a public institution) might be needed for setup or public key exchanges.
A trusted third party has an auxiliary role in the protocol, which induces costs (e.g., performing an occasional trusted setup).
If there is no trusted third party, one cannot use all the methods with ``trusted setup'' as an additional assumption in Table \ref{tab:summary_security}.
For example, NIDAR must be avoided.

\subsubsection{Questions 5: How many independent/non-colluding servers are available?}
Possible answers: Any positive integer.

\emph{Meaning:}
This question is critical to define the threat model (see Table \ref{tab:summary_security}).
For example, if only two servers are available, it is impossible to securely deploy techniques such as Blinder, which requires at least five independent servers to be secure.
We call ``independent servers'', servers controlled by independent entities (e.g., two different research institutions).
While ML researchers can own multiple physical servers, they are not independent (because they represent ``colluding'' servers sharing their information).

To filter the techniques, we need to keep all techniques whose constraint on $M$ (see Table \ref{tab:summary_security}) fits your answer.
For example, if the answer is 2 (i.e., $M=2$), one cannot use any MPC-based solutions but can use Spectrum.
Moreover, if the involved entities accept it, one can use a technique requiring less than two servers (e.g., TEE); one entity accepting to delegate the shuffling work.

\subsubsection{Questions 6: Do you need a correctness guarantee?}
Possible answers: Yes or No.

\emph{Meaning:}
Section \ref{sec:def} defines the notion of correctness and presents its importance in secure shuffling.
It can be mandatory for some applications such as DP, so this question only aims to filter based on the presence of this property (see Table \ref{tab:summary_security}).

\subsubsection{Questions 7: Are the data owners and/or the servers trusted for availability?}
Possible answers: Yes or No ($\times 2$).

\emph{Meaning:}
Similarly to the previous question, this question focus on a security property defined in Section \ref{sec:def}: disruption resistance.
If a data owner (resp. server) is untrusted for availability, it can disconnect at anytime (i.e., disrupting the protocol).
If there is no trust for availability, only the solutions with sender (resp. server) disruption resistance must be kept.

It is usual in security to trust servers for availability even if they are untrusted for data privacy.
As motivated in Section \ref{sec:def}, sender disruption resistance is particularly mandatory when the number of data owners grows since it becomes hard to assume that all will remain available.

\subsubsection{Questions 8: Do you have some application-specific requirements?}
Possible answers: None, auditability, GPU availability, etc.


\subsection{Examples}
\label{subsec:examples}
To better explain how to use these guidelines, we propose three example applications for which we choose a secure shuffler.
For each example, we give a brief presentation, the answers to the guideline questions, and the secure shuffler choice.
For the sake of simplicity, we detail \emph{only for the first example} how all the unsuitable techniques are filtered out by the guideline questions.

Each use case results into a different choice, and one even has no suitable solution.
This observation confirms no universally perfect secure shuffler exists, and gaps remain in the literature.

\subsubsection{Medical surveys}

A group of hospitals wants to deploy a privacy-preserving surveying system.
Each hospital agreed to deploy a shuffling server.
The data owners (i.e., survey participants) answer the surveys on their mobile phones and submit their answers through the shuffler.
We suppose the practitioner wants to use shuffling-based data \cite{ishai_cryptography_2006} aggregation to preserve privacy.

\emph{Answers to the guideline questions:}
(1) There are at most 10K messages (i.e., $n=10K$) per shuffle. Indirectly, we assume that it is unlikely to have millions of responses to a single medical survey.
(2) A user message is a survey response with less than 1 KB.
(3) Very constrained: the device remains online for a short period of time. Concretely, the submission should be completed a few seconds after the user presses the ``submit'' button.
(4) Yes: national health agency.
(5) There are 12 hospitals involved (i.e., $M=12$).
(6) We want correctness to ensure the correctness of the medical study.
(7) Data owners not trusted for availability. Servers trusted for availability.
(8) Specific requirements: auditability, and no hardware investment.

\emph{Secure shuffler choice.}
Mix networks are avoided because they do not guarantee correctness (Question 6).
Atom and Riffle are not resistant to sender disruption (Question 7)
All DC nets require a sender cost dependent on $n$ (Question 3).
AsynchroMix and PowerMix cannot handle so many data owners (Question 1).
Blinder requires too much computation and communication on the sender side (Question 3).
Clarion has no correctness guarantee (Question 6).
Rumor-based shuffle requires too much computation and communication on the sender side (Question 3).
NIDAR and secure-aggregation-based shufflers have no correctness guarantee (Question 6).

We end up with classic verifiable shufflers and TEE.
We argue \textbf{classic verifiable shufflers} are the adequate solution for this use case because they can be deployed instantly without investing in new hardware.
It also satisfies the auditability requirement since the (shuffle) proof transcript can be kept and audited later.

\subsubsection{AI-assisted diagnostics}

Ten hospitals want to collaborate and use their respective medical data to train a complex ML model to assist the practitioners in their diagnostics.
To train these models, the hospitals want to use shuffling-based FL.
Each hospital agrees to deploy a shuffling server.
Hence, each hospital acts simultaneously as a shuffling server and a data owner.

\emph{Answers to the guideline questions:}
(1) Less than 1K messages per round: each data owner sends one gradient per round.
(2) Messages are gradients whose size can reach few MBs.
(3) No resource constraint on the data owner side because the hospitals are the data owners.
(4) Yes: health control agency.
(5) Ten independent servers ($M=10$).
(6) Correctness guarantee wanted.
(7) Data owners and servers trusted for availability.
(8) Specific requirement: GPU available.

\emph{Secure shuffler choice.}
All MPC-based shufflers except Clarion (i.e., due to its lack of correctness) meet the expectation.
However, we argue \textbf{Blinder} is the obvious choice because it outperforms AsynchroMix and PowerMix.
Moreover, Blinder can take advantage of the available GPUs.

Finally, TEE is again an alternative, but the trust assumption must be accepted and the hospitals must choose a unique host for the secure enclave.

\subsubsection{Smart keyboard application}
This last use case was used as motivation in the paper that sparked the interest in FL \cite{konecny_federated_2016}: a Google-like company develops a smartphone keyboard application and wants to improve the next-word recommendations.
Each smartphone locally trains a model based on the application usage, and a central server securely aggregates the local models to produce a global recommendation model.
We want a single-server secure shuffler to perform the secure aggregation step because the company does not want to involved any third party.
Google-like services have millions (or even billions) of users, so the scalability expectations are high.

\emph{Answers to the guideline questions:}
(1) Tens of millions of data owners.
(2) Messages are gradients whose size can reach a few MBs.
(3) Very constrained.
(4) No trusted third party.
(5) One server available: one company = one independent server.
(6) No correctness guarantee needed.
(7) Data owners not trusted for availability. Server trusted for availability.
(8) Specific requirement: the company does not want to rely on an open-source community such as TOR network due to the economic interest of the application.

\emph{Secure shuffler choice.}
There is \textbf{no solution} satisfying all the requirements.
The best solutions are TEE-based shufflers, but Prochlo suffers from side-channel attacks, and DBucket has a non-linear complexity (i.e., $O(n\log ln)$) that could be acceptable in practice.
However, TEE might be unsatisfactory since it requires trusting an independent hardware manufacturer.

%% file: sec/takeaways.tex
\section{Perspectives on the literature}
\label{sec:takeaways}

This section introduces several general perspectives on the literature.
The subsections are independent of each other and discuss misconceptions or subtle details that are often overlooked in the literature.
Moreover, these perspectives enable to identify several promising future works.

\subsection{Preventing metadata leakage is unnecessary}
\label{subsec:ultimate_anon}

Our work focused on the most conventional definition of sender anonymity, but some works also proposed stronger anonymity notions than Definition \ref{def:anonymity_vs}: referred to as ultimate anonymity \cite{gelernter_limits_2013} or perfect anonymity \cite{backes_anoa_2013}.
This stronger anonymity forbids the metadata leakage mentioned in Section \ref{sec:def}: the size and number of sent messages.

From a theoretical point of view, it combines sender anonymity with a notion called ``unobservability''.
Unobservability ensures that the observation of the traffic leaks absolutely no information.
Gelernter and Herzberg \cite{gelernter_limits_2013} show that ultimate anonymity is inefficient because each sender must generate a large amount of cover traffic to prevent metadata leakage.
Indeed, each sender must send $O(n)$ data to be a possible sender of the whole batch, even if the sender has small or no data to send.
This result explains why Dissent (the only secure shuffler satisfying ultimate anonymity) is the least scalable technique.

While leakage sounds like a side effect to avoid at any cost, this metadata leakage is acceptable in many applications, especially in privacy-preserving computations.

On the one hand, privacy-preserving computations usually involve messages containing fixed-sized vectors or atomic elements (e.g., floating-point numbers).
Thus, they are all the same size.
On the other hand, many privacy-preserving computations define the number of messages sent by each data owner independently of their private data.
For example, in shuffling-based secure aggregation \cite{ishai_cryptography_2006}, each data owner must send the same number of messages.
To sum up, (shuffling-based) privacy-preserving computations do not require anonymity stronger than the sender anonymity defined in Section \ref{sec:def} because the message metadata is often fixed and known publicly.

\subsection{Imperfect shuffling/anonymity can be acceptable}
\label{subsec:imperfect_shuffle}
Shuffling-based private computations (including in DP) usually assume a perfect shuffle (i.e., uniform anonymity distribution as defined in Section \ref{sec:anonymity}).
However, \citet{ishai_cryptography_2006} claimed that only a ``sufficient uncertainty'' was necessary; i.e., imperfect shuffle could be acceptable.
Recently, \citet{balle_amplification_2023} showed that this intuition was true as their results holding also for an imperfect shuffle.

These results should incentivize DP researchers to build DP algorithms robust to ``slightly imperfect'' shuffling and security researcher to build protocols leveraging this imperfection.
Nevertheless, further theoretical work is also necessary to standardize this notion of imperfect shuffling.

\subsection{Avoid modifying the shuffler functionality}
Even though shuffle DP works agree on the intuitive definition of a shuffler, there is no consensus regarding the exact capability of a shuffler.
Most papers \cite{cheu_distributed_2019,cheu_differential_2021} rely on the most basic definition as we do (i.e., the shuffler simply outputs a permutation), but some papers \cite{yang_adastopk_2023} complete this functionality with other simple tasks such as adding unique identifiers to the messages.

We argue that such modifications should be avoided because these modified functionalities cannot be implemented naively with all existing secure shufflers.
Thus, DP researchers should stick to the most basic and generic shuffler functionality to guarantee that the shuffling functionality can be implemented securely.
We recommend DP researchers to adopt a ``generic-shuffler'' approach, assuming a generic behavior as defined in Definition \ref{def:shuffler}.
Such an approach would guarantee that their DP algorithm can be implemented using any of the shufflers listed in our survey.
While most works already adopt this approach, our findings highlight its importance.

One may find our recommendation contradictory with the acceptability of an imperfect shuffle, but the imperfect shuffle does not modify the functionality (i.e., the shuffler still outputs a permutation), it simply modifies the security definition.

Finally, this generic approach should also be completed with some security discussions.
Indeed, shuffle DP works assume a perfect shuffler and do not discuss the consequences of a flawed shuffler.
For example, what are the consequences if the shuffle output was partially censored (i.e., no correctness)?
Such security discussions are absent from existing DP works, but would provide a much better understanding of the privacy limitations.
Using the security definitions proposed in Section \ref{sec:def}, DP researchers would have a standardized set of properties to discuss.

\subsection{Information-theoretic Shuffle DP is impractical}
\label{subsec:comp-shuffle-dp}
Computational DP \cite{beimel_distributed_2008,mironov_computational_2009} is a relaxed variant of DP where the DP guarantees hold against computationally-bounded adversaries.
This definition is opposed to ``information-theoretic'' DP (i.e., the initial DP definition proposed by Dwork \cite{dwork_differential_2006}), where the guarantees also hold against computationally unbounded adversaries.
This separation is classic in security, where we oppose computational security to information-theoretic security.

A DP algorithm is ``computationally differentially private'' if one of its components is only computationally secure.
For example, as mentioned in \cite{bun_separating_2016}, information-theoretic DP does not permit using cryptographically secure pseudo-random generators in place of perfect randomness.

This constraint (i.e., no computationally secure component) is problematic for Shuffle DP.
As mentioned in Section \ref{sec:comparison}, all secure shufflers assume the existence of a secure communication channel between the senders and the shuffler.
However, Maurer \cite{maurer_information-theoretic_1999} showed that information-theoretically secure communication channels are impractical.
Therefore, we should only assume computationally secure communication channels in the context of shuffle DP.
Hence, \textbf{only computational shuffle DP can be implemented efficiently}.

Relying on ``information-theoretic'' shuffle DP is then impractical because any real world deployment requires a computational hardness assumption.
Thus, shuffle DP applications should take this mandatory assumption into consideration in their analysis.
Such consideration should incentivize more ``computational'' DP results; known to require less noise than information-theoretic DP \cite{mironov_computational_2009}.

This conclusion holds for all DP protocols assuming the existence of a secure communication channel between two parties.

\subsection{Malicious data owners are a serious threat}
\label{subsec:malicious-sender}
Our security properties hold against an unbounded number of malicious data owners.
Anonymous communications systems commonly assume a ``sufficiently high'' number of honest senders, as anonymity holds within the set of honest senders.
For example, Langowski et al. \cite{langowski_trellis_2023} highlighted this limitation when presenting their anonymous broadcaster.

This vague assumption is acceptable because the number of malicious data owners has no influence on the security properties of a secure shuffler.
However, this number has an influence on the privacy and correctness of the overall shuffling-based private computation.

\paragraph{Threat to privacy}
First, malicious data owners can decrease the privacy guarantees in Shuffle DP computations because the DP noise is proportional to the number of honest data owners.
Indeed, if a malicious data owners does not generate a noise following the agreed protocol, the output result will provide a weaker privacy than expected.
Hence, knowing at least a lower bound on the number of honest data owners is necessary.

To verify that malicious data owners add an appropriate noise, Biswas and Cormode \cite{biswas_verifiable_2022} introduced the notion of verifiable DP (generalized to Shuffle DP by \citet{bontekoe_efficient_2024}); leveraging zero-knowledge proofs to verify the DP noise generation and addition.

\paragraph{Threat to correctness}
Malicious data owners also threaten the computation correctness because they can bias the final results via data poisoning attacks.
For example, in FL, this takes the form of model poisoning attacks \cite{cao_data_2021,carlini_poisoning_2021,schuster_you_2021} in which attackers submitting local models that would introduce undesired behaviors in the global model.
Such poisoning can have serious consequences as it can, for example, add a backdoor to an ML model \cite{schuster_you_2021}.

Norm bounding is a popular mitigation technique \cite{shejwalkar_back_2021}.
This simple mitigation has been incorporated into secure aggregation protocols using ZKPs \cite{bell_acorn_2022,chowdhury_eiffel_2022}.

However, no mitigation compatible with shuffling-based protocols has been yet described.
This verification is trivial when each data owner sends a single message because the verification can be done on the shuffle output.
However, in multi-message Shuffle DP \cite{cheu_differential_2021} or shuffling-based secure aggregation \cite{ishai_cryptography_2006}, each data owner splits its private value into several messages.
A zero-knowledge verification is necessary to verify claims about the initial private value while protecting anonymity.

\subsection{Trusted hardware is not a silver bullet}
\label{subsec:tee-strong-assumptions}

In theory, trusted hardware (i.e., TEE) is an appealing assumption that can solve privacy issues: the protocol is secure as long as the data goes through the trusted hardware (even with a malicious server owner).
Several papers promoted TEE in FL \cite{bittau_prochlo_2017,boutet_mixnn_2021,zhang_shufflefl_2021,zhao_sear_2022} motivated by its performances and (theoretical) security.
However, one must also understand the strength of the assumption required by TEEs: the data owner fully trusts the manufacturer.
TEE also requires trusting the protocol implementation because there exist other flaws: side-channel attacks \cite{brasser_software_2017,wang_leaky_2017}.

Sasy et al. \cite{sasy_fast_2022} presented an excellent example of the risk of TEE.
While some shuffle DP papers used Prochlo \cite{bittau_prochlo_2017} (a TEE-based secure shuffler) as motivation, Sasy et al. \cite{sasy_fast_2022} argued that recent attacks against TEE make Prochlo insecure.
Hence, even if Prochlo seemed at first totally secure, it turned out to be insecure in practice due to TEE vulnerabilities.

Finally, as TEE relies on particular hardware, one requires compatible hardware.
It is a solution that cannot be deployed anywhere easily; it requires a dedicated hardware investment.